\documentclass[sigconf, 10pt]{acmart}

\usepackage{tikz}
\usepackage{amsmath}
\usepackage{amsthm}

\usepackage{filecontents}
\usepackage{diagbox}
\usepackage[english]{babel}
\usepackage{blindtext}
\usepackage{xcolor}
\usepackage{gincltex}
\usepackage{pgfplots, siunitx, tikzscale}
\usepackage{graphicx}
\usepackage{subfigure}
\usepackage{pifont}
\usetikzlibrary{pgfplots.statistics}
\usepackage{float}
\usepackage{tcolorbox}
\usepackage { bb m}
\usepackage{enumitem}
\usepackage{booktabs}
\usepackage{makecell}
\usepackage {pifont}
\usepackage{amssymb}
\usepackage{algorithm}
\usepackage{algpseudocode}
\usepackage{tabularx}
\usepackage{multirow}
\usepackage{multicol}
\usepackage[font=bf, size=small]{caption}
\newcommand{\name}{KVCodec\xspace}
\newcommand{\hide}[1] {}

\definecolor{mypink1}{rgb}{0.858, 0.188, 0.478}

\vspace{0.4em}
\newcommand{\ie}{\textit{i.e.}\xspace}
\newcommand{\eg}{\textit{e.g.}\xspace}

\usepackage{titlesec}
\titleformat{\subsection}
  {\normalfont\fontsize{10.5}{12}\bfseries}{\thesubsection}{1em}{}
\titleformat{\subsubsection}[runin]{\normalfont\bfseries}{\thesubsubsection}{1em}{}
\titlespacing*{\subsubsection}{0pt}{\dimexpr 1ex minus 0.7ex}{\dimexpr 1ex plus 0.7ex}

\title{\huge Efficient Remote KV Cache Reuse with GPU-native Video Codec}

\author{
{Liang Mi$^{1}$\footnotemark[1]\footnotemark[2]\;Weijun Wang$^2$\footnotemark[1]\;Jinghan Chen$^1$\;Ting Cao$^{2}$\footnotemark[3]\;Haipeng Dai$^1$\footnotemark[3]\;Yunxin Liu$^{2}$}\\
$^1$ Nanjing University
$^2$Institute for AI Industry Research (AIR), Tsinghua University
} 

\renewcommand\footnotetextcopyrightpermission[1]{} 
\setcopyright{none}

\acmDOI{}

\acmISBN{}

\acmPrice{}
 
\begin{document}



\begin{abstract}
Remote KV cache reuse fetches KV cache for identical contexts from remote storage, avoiding recomputation, accelerating LLM inference. 
While it excels in high-speed networks, its performance degrades significantly in bandwidth-limited scenarios.
Recent studies address this by transmitting KV caches in compressed form, but the associated heavyweight decompression counteracts the KV reuse benefits.
In this paper, we propose an \textit{efficient} and \textit{widely deployable} remote KV cache reuse solution that leverages \textit{GPU-native video codecs}.
Our system, \name, enables effective KV cache coding with two techniques. 
The \textit{codec-friendly tensor layout} compresses the KV cache in a highly compact video format, enabling fast transmission. 
The \textit{efficient KV fetcher} orchestrates the transmission, decoding, and restoration of compressed KV caches in an efficient pipelined manner, eliminating resource contention, masking network fluctuations, and achieving minimum time-to-first-token (TTFT).
We prototype \name on diverse GPUs from high- to low-end.
Experiments reveal that it reduces TTFT by up to 3.51$\times$ while maintaining lossless accuracy, compared to SOTA methods. 
\end{abstract}

\maketitle

\renewcommand{\thefootnote}{\fnsymbol{footnote}} 

\footnotetext[1]{Liang Mi and Weijun Wang contributed equally to this work.} 
\footnotetext[2]{This work was done while Liang Mi interns at the Institute for AI Industry
Research (AIR), Tsinghua University.} 
\footnotetext[3]{Corresponding author: Ting Cao and Haipeng Dai.}

\section{Introduction}
With context window trending to millions of tokens, Large Language Models (LLMs) have demonstrated superiority in many modern services. 
By ingesting informative contexts, including system prompts~\cite{SystemPrompt}, retrieved documents~\cite{comanici2025gemini25pushingfrontier, Claude}, and interaction histories~\cite{beltagy2020longformer, borgeaud2022improving, zhuang2025self, wu2025agentic}, together with request queries, LLMs can generate high-quality outputs powering diverse applications, from 
code generation~\cite{chen2021evaluating, guo2024deepseek, roziere2023code} to 
agentic workflows~\cite{hong2023metagpt, wang2025agentspec, yang2024swe}.
For efficient serving, storing the intermediate states, \textit{KV cache}, during LLM inference has become a de facto choice.



\textit{KV cache reuse} further amortizes the computational costs across multiple requests. 
It persists KV caches in the first inference and reuses them when future requests share common prefixes.
Most LLM inference systems~\cite{ye2024chunkattention, gim2023prompt, kwon2023efficient, zheng2024sglang, jin2024ragcache, qin2025mooncake, lmcache, aibrix} today integrate KV cache reuse as their key feature.
Their consensus is that KV caches are frequently reused. 
Mooncake~\cite{qin2025mooncake} claims that 50\% of KV caches will be reused in KiMi's real-world workloads, while LMCache~\cite{lmcache} believes this number is higher. 
Despite this benefit, the storage cost of KV caches can be huge~\cite{kwon2023efficient, xiang2025shadowserve}. 
Storing 80K-token KV caches (\eg, Amazon's annual report) of a medium-level 34B model can consume up to 19GB of storage~\cite{liu2024cachegen}.
For this reason, hosting all reusable KV caches locally is impractical.

\begin{figure}[t]
\centering
    \setlength{\abovecaptionskip}{5pt}
    \includegraphics[width=1\linewidth]{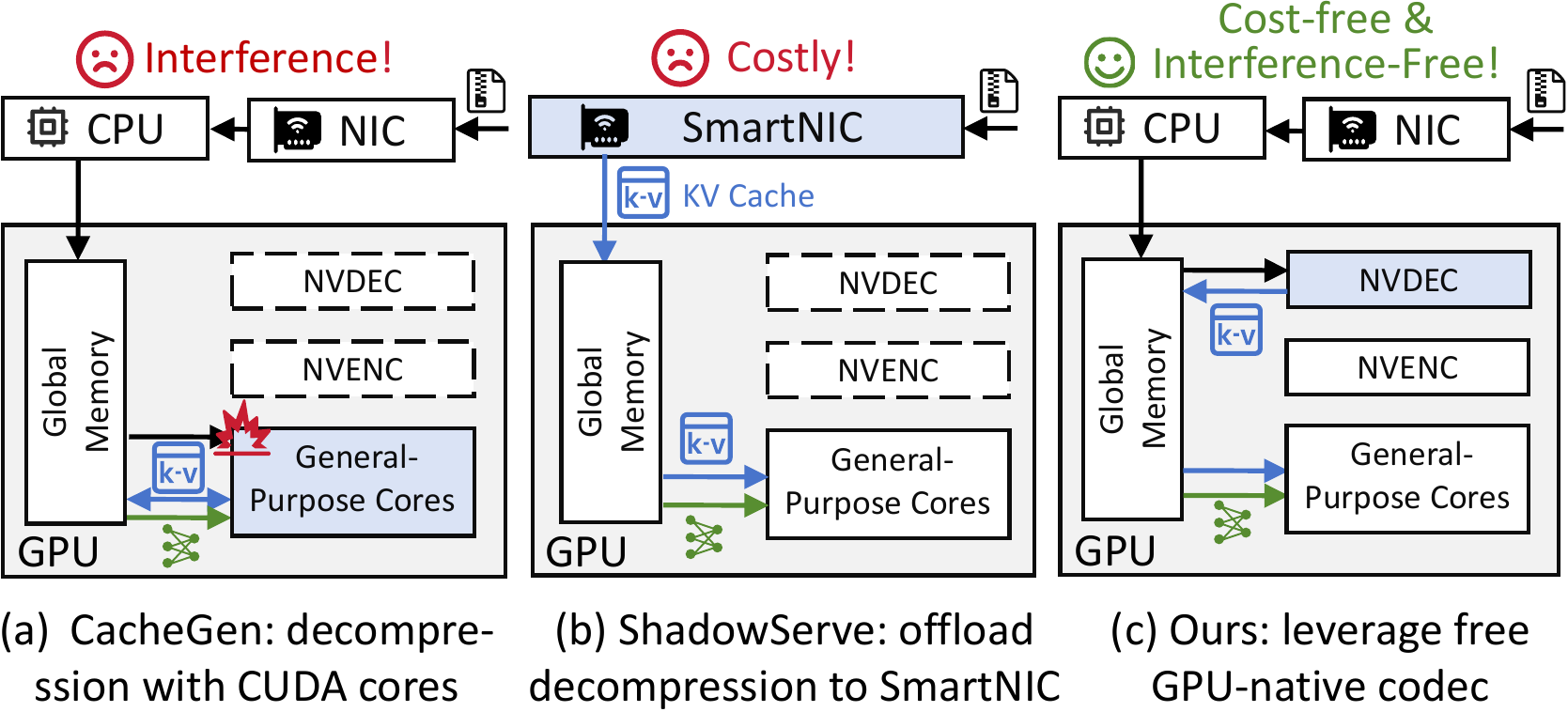}
    \caption{Solutions of remote KV cache reuse. Our \name exploits GPU-native video codecs, delivering the best cost-efficiency and system performance.
    }
    \label{fig:NVCODEC}
    \vspace{-1em}
\end{figure}


To tackle this, KV cache reuse uses external storage and fetches them on demand from remote resources. 
Mooncake abstracts all GPU memory, host memory, and disks within GPU clusters into a disaggregated KV cache pool, 
and LMCache leverages dedicated storage servers.
In such distributed setups, KV caches must be transmitted over the network.
Although high-speed interconnections (\eg, RDMA) render transmission latency negligible~\cite{qin2025mooncake, zhong2024distserve, strati2024DejaVu, lin2024infinite}, economic considerations drive modern LLM services to deploy on mid-range GPUs typically paired with only constrained bandwidths, tens of Gbps or less ~\cite{mao2025skyserve, miao2024spotserve, xiang2025shadowserve, mei2025helix, Dynamo, jiangdemystifying}. 


Recent studies~\cite{liu2024cachegen, xiang2025shadowserve} mitigate this networking bottleneck by transmitting KV caches in compressed form rather than raw tensors. 
It indeed yields bandwidth savings, but the associated decompression latency lies on the critical path of LLM inference, counteracting the network benefits.
CacheGen~\cite{liu2024cachegen} customizes a CUDA kernel to accelerate decompression, yet it competes for GPU resources with LLM inference engines, as the blue box shown in Fig.\ref{fig:NVCODEC}(a), severely degrading LLM inference performance (\S\ref{sec:Limitation}). 
ShadowServe~\cite{xiang2025shadowserve} achieves interference-free inference by offloading decompression to SmartNIC, as shown in Fig.\ref{fig:NVCODEC}(b); but, the prohibitive hardware cost hinders its widespread adoption. 
We believe KV cache compression is promising, but requires a cost-efficient redesign to be widely used.


Modern GPUs are typically equipped with dedicated hardware for efficient video coding.
Although initially designed for media processing, their independent on-chip computing and storage units offer a great opportunity to address the above limitations (\S\ref{sec:OurChoice}).
More importantly, as dashed boxes shown in Fig.\ref{fig:NVCODEC}, they are completely idle throughout the LLM inference.
Motivated by this, we ask: \textit{Can we implement a KV cache codec with GPU-native video codecs, enabling efficient remote KV cache reuse?} 
A concurrent study, llm.265~\cite{xu2025llm265}, attempts to compress KV caches by video coding, but cannot be the answer to this question.
It provides very limited compression gain and lacks system co-design with inference engines (\S\ref{sec:challenges}).
To be the answer, we must tackle two challenges.

First,  to minimize transmission delay, efficient remote KV cache reuse necessitates highly compact compression.
During compression, it must map KV tensors to a video format and encode them into bitstreams using video encoders.
However, this is non-trivial.
A successful compression strategy must comprehensively explore and judiciously exploit both the distribution of K and V values and the characteristics of the video coding, to maximize the compression ratio without compromising LLM generation quality.
Blindly representing the KV tensor as individual video frames and directly encoding them into the bitstream, like llm.265~\cite{xu2025llm265}, results in both suboptimal compression ratios and generation quality (\S\ref{sec:challenges}).

Second, to minimize the Time to First Token, TTFT, only compact compression is not enough.
It just accelerates the transmission of single video chunks, but efficient remote KV reuse still requires considerate scheduling as well as rapid and interference-free decoding and restoration\footnote{This paper follows the same setting as prior studies~\cite{liu2024cachegen, xiang2025shadowserve, qin2025mooncake}, where the KV caches are chunked and compressed in advance, and stored at remote storage nodes. So remote KV reuse (a.k.a., KV fetching) consists of KV transmission, decoding, and restoration.}.
Existing KV cache reuse systems~\cite{lmcache, qin2025mooncake} hardly meet these requirements.
First, their schedulers indiscriminately batch all arriving requests, leading to non-reuse requests (requests without remote KV cache reuse) being blocked by the fetching stage of requests with remote reuse.
Second, their bulk KV restoration consumes excessive memory from LLM inference engines, delaying the TTFT. 
And last, their ill-considered design hardly copes with the networking jitter (\S\ref{sec:challenges}).
In this paper, we present \name, an efficient remote KV cache reuse module powered by GPU-native video codecs, to address the challenges outlined above.

To address the first challenge, we propose a \textit{codec-friendly tensor layout} that enables a tenfold compression ratio without a drop in LLM inference accuracy.
Its core idea is to skip the lossy Discrete Cosine Transform (DCT) and quantization steps of video encoding, and fully utilize the lossless intra- and inter-frame redundancy elimination capability.
To achieve this, we conduct an in-depth analysis of the characteristics of video coding, the distribution of KV values, the properties of LLM architectures, and finally obtain the layout principle: slice KV tensors along the token dimension, scatter resulting tensors over continuous frames, and encode them in multiple-resolution versions (more in \S \ref{sec:KVCompress}).


To tackle the second challenge, we propose an \textit{efficient remote KV fetcher} (\S\ref{sec:Fetching}) including three key techniques. 
The fetching-aware scheduler (\S\ref{sec:Scheduler}) discriminates the requests w./w.o KV reuses and isolates the KV fetching in the background, preventing blocking the inference of non-reuse requests. 
The adaptive-resolution KV fetching (\S\ref{sec:decompression}) adjusts the video resolution to tune the video size, enabling efficient KV fetching across varying network bandwidths. 
The frame-wise tensor restoration minimizes the memory cost of restoring decoded frames to original KV tensors, without impacting LLM inference engines.

We summarize our key contributions as follows:

\scalebox{0.8}{$\bullet$} To the best of our knowledge, we are the first to identify the opportunity of GPU-native video codecs for remote KV cache reuse to accelerate LLM inference.

\scalebox{0.8}{$\bullet$} We prototype \name, an efficient remote KV cache reuse system powered by GPU-native video codecs. 
It involves two core techniques, codec-friendly tensor layout and efficient remote KV cache fetching.

\scalebox{0.8}{$\bullet$} We implement \name and conduct evaluations on three GPUs from high- to low-end and three different size models from 7B to 70B over 1-40Gbps bandwidths. 
Experimental results show that \name achieves 1.52-3.51$\times$ TTFT reduction compared to SOTA methods, while ensuring high accuracy and interference-free non-reuse requests inference. 


\textbf{This work does not raise any ethical issues.}

\section{Motivation and Challenges}

\subsection{Preliminary of Remote KV Cache Reuse}\label{sec:Preliminary}
\noindent
LLM inference contains two phases.
The prefilling phase processes the entire input tokens simultaneously to generate the first output token, while the decoding phase generates the subsequent ones in an autoregressive manner. 
For efficiency, LLM inference engines store intermediate tensors, \ie, the KV caches produced by each attention layer, to eliminate the redundant computation of autoregressive generation.
Although KV cache significantly speeds up decoding, it provides no acceleration for the prefilling phase.


\textbf{KV cache reuse} operates as a storage-for-computation trade-off, substantially reducing prefilling latency. 
Its core principle is that KV caches generated by one request can be reused by subsequent requests sharing a common sequence prefix. 
This pattern is highly prevalent across modern LLM applications.
For example, to maintain consistency, chatbots~\cite{gao2024cost, yu2025stateful, gao2025fast, jeong2025accelerating} feed conversation histories into the LLM at each chat round;
in multi-agent code debuggers~\cite {cursor, qian2023communicative, yang2024swe}, web-search agents frequently reread buggy code from coding agents to find related posts on the Internet; 
and Vision-Language-Action (VLA) models~\cite{lin2025onetwovla, shi2025memoryvla, lei2025robomemory} in embodied AI continuously recall prior observations and actions to boost the quality of the next action.
Since this repeated content can be leveraged after the first processing, LLM inference engines \cite{jin2024ragcache, zheng2024sglang} persist their KV caches, rather than immediately freeing them, to facilitate future reuse.

\textbf{Remote KV cache reuse} significantly extends the capacity of reusable KV cache by retrieving them from remote resources.
In large-scale LLM services~\cite{qin2025mooncake, liu2024cachegen, fu2025cache, aibrix, xiang2025shadowserve}, since the limited capacity of single node and the strategies of load-aware~\cite{cao2025locality, hu2024memserve} and fairness-aware~\cite{srivatsa2024preble} request dispatching, reusable KV caches are typically scattered across distant nodes.
To reuse these KV caches in this distributed environment, LLM inference engines must fetch them from the source nodes via the network interconnect.
However, remote KV cache reuse is only beneficial when the KV cache fetching speed exceeds the latency of recomputation.
Early LLM serving systems~\cite{qin2025mooncake, strati2024DejaVu, lin2024infinite} could easily achieve this benefit, as they were commonly deployed in High-Performance Computing (HPC) centers with >100Gbps RDMA interconnects.
In contrast, for cost efficiency, modern serving systems~\cite{cao2025moe, griggs2024m, mei2025helix, xiang2025shadowserve} tend to be deployed on mid-range or even low-end GPUs, which communicate over bandwidth-limited networks with tens of Gbps or less.
Some other systems~\cite{aibrix, Dynamo} rent dedicated storage servers to scale the capacity of reusable KV caches, yet their performance is still limited by constrained bandwidth (19 Gbps on AWS~\cite{awsprice}).

\begin{figure}[t]
\centering
\setlength{\abovecaptionskip}{4pt}
    \includegraphics[width=0.45\textwidth]{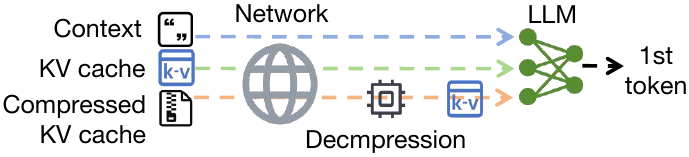}
    \caption{Current three prefilling types: full prefill, raw KV reuse, compressed KV reuse.
    Their time costs are: prefill, transmission+prefill, transmission+decompression+prefill.}
    \label{fig:RemoreKVReuse}
\end{figure}
\begin{figure}[t]
\centering
\setlength{\abovecaptionskip}{4pt}
     \centering
    \includegraphics[width=0.7\linewidth]{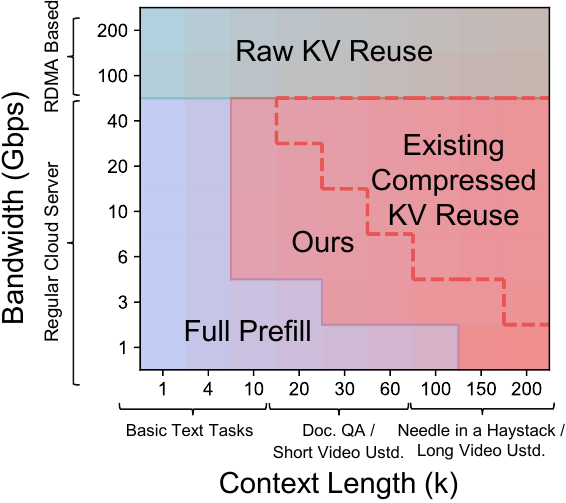}
    \caption{``Winning areas'' of three prefilling types under various bandwidths and context lengths. \name significantly extends the applicable scope of compressed KV reuse.
   }
      \label{fig:Position}
    \vspace{-1em}
\end{figure}

\subsection{Limitations of Existing Remote KV Cache Reuse Systems}\label{sec:Limitation}
To mitigate the network bottleneck, SOTA systems~\cite{liu2024cachegen, xiang2025shadowserve} transmit KV caches in compressed forms. 
While theoretically beneficial, our empirical analysis reveals that these bandwidth savings yield only a limited end-to-end latency reduction (as elaborated shortly).
To explore the reason, we benchmark the TTFT of three approaches, as shown in Fig.\ref{fig:RemoreKVReuse}: (1) \textit{Full Prefill} (baseline without KV reuse), (2) \textit{Raw KV Reuse} (\eg, Mooncake~\cite{qin2025mooncake}, AIBrix~\cite{aibrix}), and (3) \textit{Compressed KV Reuse} (\eg, CacheGen~\cite{liu2024cachegen}, ShadowServe~\cite{xiang2025shadowserve})\footnote{These transmission-oriented methods leverage aggressive arithmetic coding to encode KV caches into bitstreams, which are orthogonal to prior KV cache compression methods (\eg, pruning~\cite{pan2024llmlingua, zhang2023h2o} or quantization~\cite{liu2024kivi, hooper2024kvquant}) that keeps tensor formats for GPUs directly use.}, as illustrated in Fig.\ref{fig:RemoreKVReuse}. 
The evaluation is conducted on a 200k-context-window LWM-7B~\cite{ai2025yiopenfoundationmodels} model served by vLLM~\cite{kwon2023efficient} on 2 NVIDIA H20 GPUs. 
Requests arriving follow the real-world trace~\cite{kvcaches}, and network bandwidth is regulated from 1 to 40 Gbps over TCP and 100/200 Gbps over RDMA.
As shown in Fig.~\ref{fig:Position}, the ``winning area'' of the existing compressed KV reuse solution (dashed box) is surprisingly small for three reasons.

\begin{figure}[t]
\centering
\setlength{\abovecaptionskip}{5pt}
  \begin{minipage}[b]{0.48\linewidth}
  \centering
   {\includegraphics[width=1\linewidth]{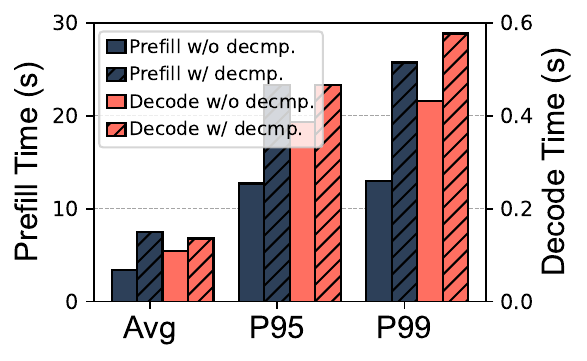}}
    \caption{Concurrent LLM inference and KV decompression cause extra delay.}
    \label{fig:Impacts}
    \end{minipage}
\hfill
  \begin{minipage}[b]{0.48\linewidth}
  \centering
    {\includegraphics[width=1\linewidth]{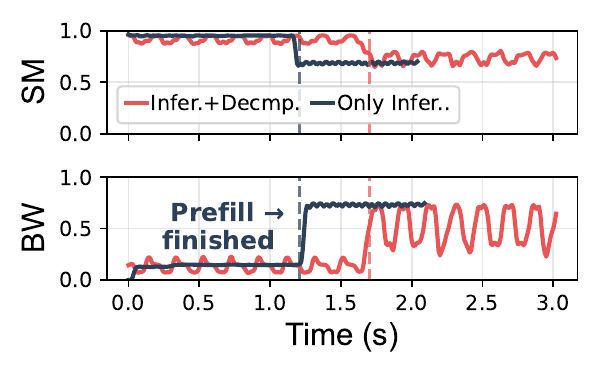}}
    \caption{Kernel switch yields SM underutilization and memory I/O contention.}
      \label{fig:Utilization}
  \end{minipage}
  \vspace{-1em}
\end{figure}

\textbf{Limited TTFT gains due to suboptimal compression.}
Both ShadowServe and CacheGen treat KV tensors as generic byte streams and compress them with arithmetic coding, resulting in a low compression ratio.
Such a compression strategy entirely ignores the unique distribution of K and V data (\S\ref{sec:KVCompress}), rendering the reduced transmission time unable to amortize the substantial decompression delay and thus poor TTFT in most bandwidth-context-length scenarios.

\begin{figure}[t]
\centering
\setlength{\abovecaptionskip}{5pt}
     \centering
    \includegraphics[width=0.3\textwidth]{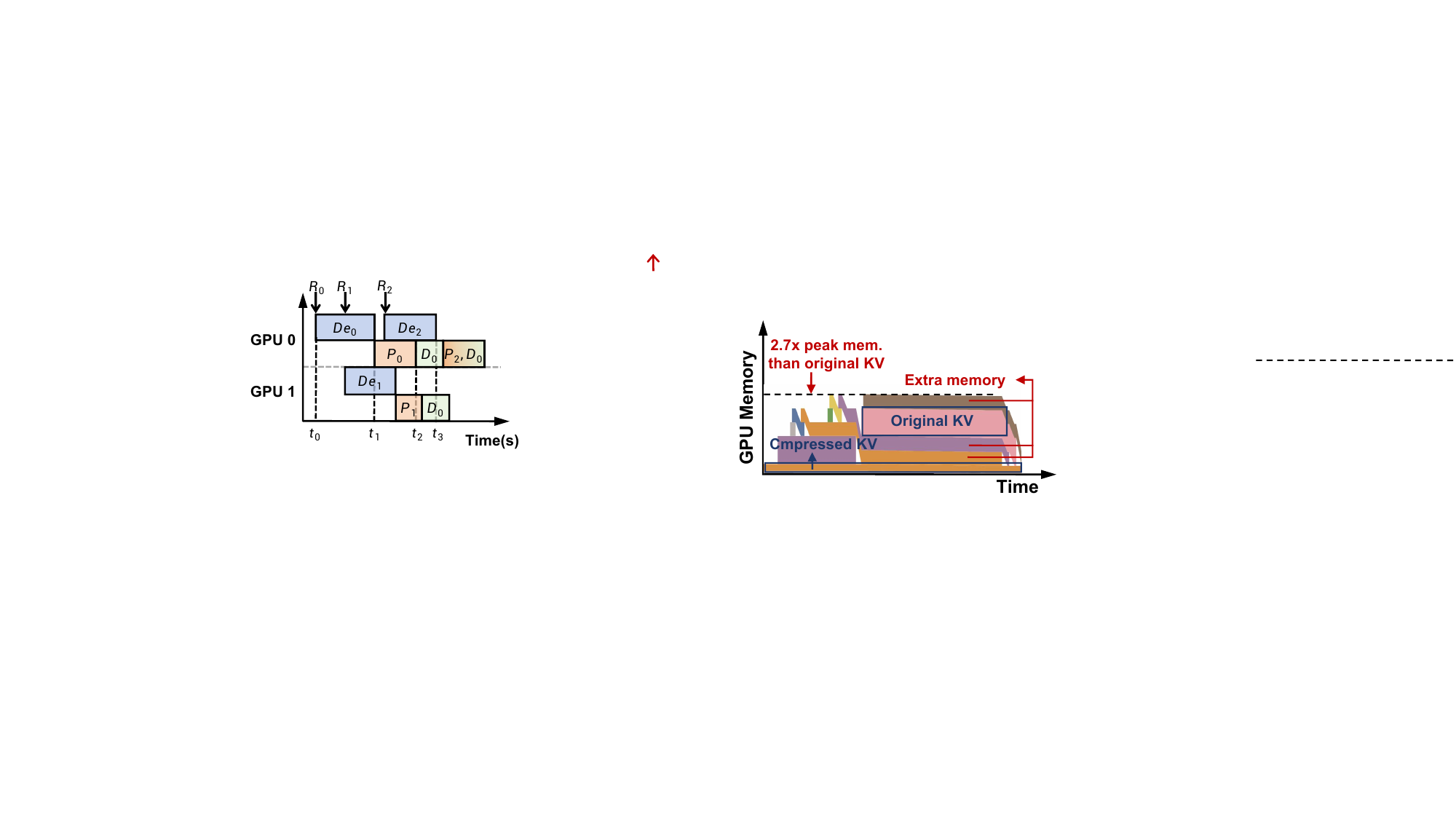}
    \caption{Peak memory of decompression in CacheGen is 2.7$\times$ that of the raw KV cache.}
    \label{fig:PeakDecomp}
    \vspace{-1.5em}
\end{figure}

\textbf{Severe resource contention of CUDA-based decompression.}
CacheGen customizes a CUDA kernel to accelerate the decompression of KV caches.
It truly decreases decompression time but causes a significant LLM inference delay, resulting in a 50\% increase in prefilling time and a 20\% increase in decoding time, as shown in Fig.~\ref{fig:Impacts}.
This stems from two factors.
({\romannumeral 1}) \textit{Compute preemption:}
Fig.\ref{fig:Utilization} profiles the Streaming Multiprocessors (SM) and GPU I/O bandwidth utilization of standalone LLM inference versus concurrent with decompression.
Compared to standalone (blue curve), concurrency triggers frequent kernel context switching (the fluctuation of the red curve), leading to SM underutilization, high memory I/O contention, and thus delaying the LLM inference.
({\romannumeral 2}) \textit{Memory bloat:} 
CacheGen's decompression seizes a considerable amount of GPU memory from LLM inference engines.
As shown in Fig.\ref{fig:PeakDecomp}, it pre-allocates 5.5GB of GPU memory, $2.7\times$ larger than the original KV cache, to decompress only 4K tokens. 
This ill-considered memory management limits the batch size of LLM inference, further increasing the average inference time.

\textbf{Prohibitive deployment costs of SmartNIC-based decompression.}
To avoid GPU contention, ShadowServe offloads KV cache decompression to SmartNICs. 
It isolates the decompression from LLM inference but introduces significant deployment costs (>\$3000 of each NVIDIA BlueField NIC) and hardware dependencies. 
These constraints limit the scalability and adoption of ShadowServe.
In line with the current trend of deploying LLM inference systems on economical platforms, this paper aims to develop a cost-efficient and easy-deployed solution. 

\begin{figure}[t]
\centering
\setlength{\abovecaptionskip}{5pt}
\includegraphics[width=0.9\linewidth]{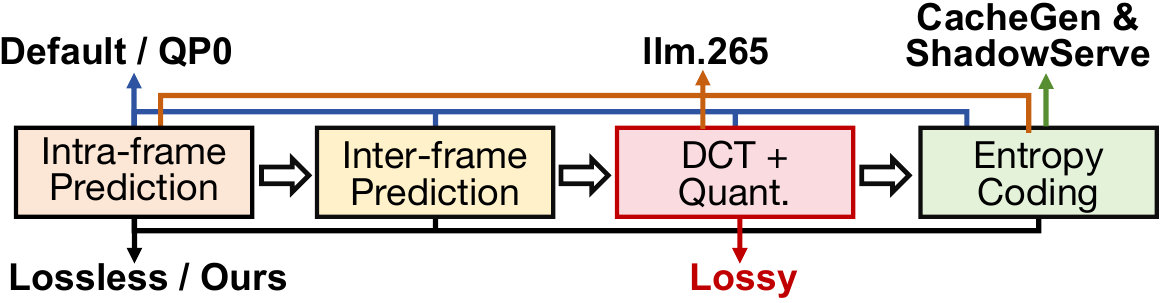}
    \caption{Standard H.265 encoding pipeline. Default and QP0 apply all steps to KV tensors while Lossless skips lossy steps. llm.265~\cite{xu2025llm265} skips inter-frame prediction. CacheGen~\cite{liu2024cachegen} and ShadowServe~\cite{xiang2025shadowserve} only utilize arithmetic coding.}
    \label{fig:encode_flow}
    \vspace{-1em}
\end{figure}

\subsection{
Opportunity of GPU-native Video Codec 
}\label{sec:OurChoice}
%
Modern GPUs are typically equipped with dedicated ASIC units, \eg, NVIDIA's NVENC/NVDEC~\cite{nvidiagpu}, AMD's AMF~\cite{amdgpu}, and Intel's QSV~\cite{intelgpu}, for efficient video coding.
Despite being designed for video, their on-chip computing and storage resources, which are totally independent of general-purpose computing units, offer us a good chance.
More importantly, current LLM inference engines (\eg, vLLM~\cite{kwon2023efficient}, SGLang~\cite{zheng2024sglang}) completely ignore these resources, leaving them fully idle during inference, as illustrated in Fig.~\ref{fig:NVCODEC}.  

\textbf{Offloading KV cache compression and decompression to GPU-native video codecs} opens up a new design space for remote KV cache reuse.
It presents a valuable opportunity to address the limitations of existing methods. 
({\romannumeral 1}) Video codecs exploit spatial and temporal redundancies of images (tensors), offering the potential to achieve superior compression ratios beyond generic AC methods.
({\romannumeral 2}) Offloading to independent hardware isolates the workload, effectively eliminating the compute preemption and memory bloat, enabling interference-free simultaneous inference and KV decompression.
({\romannumeral 3}) As ubiquitous components on modern GPUs, this ASIC provides a scalable solution with zero additional hardware cost.
While llm.265~\cite{xu2025llm265}, a concurrent study, also attempts to compress KV caches with video coding, it fails to fully exploit the compression potential of video codecs and lacks co-design with LLM inference engines (\S\ref{sec:challenges}). 
\subsection{Challenges of Compressed KV Streaming with GPU-native Video Codec}\label{sec:challenges}
Building an efficient remote KV cache reuse system using GPU-native codec hardware involves two challenges.

\begin{figure}[t]
\centering
\setlength{\abovecaptionskip}{5pt}
  \begin{minipage}[t]{0.49\linewidth}
    \raisebox{0.15cm}{\includegraphics[width=1\linewidth]{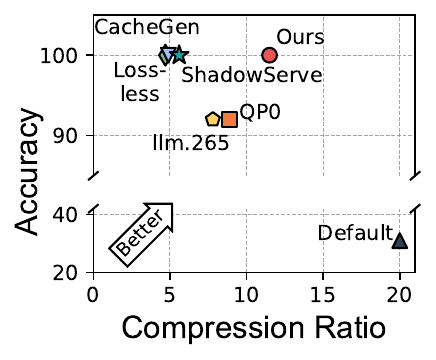}}
    \caption{Current solutions yield an unsatisfactory tradeoff between accuracy and compression.}
      \label{fig:AccComp}
\end{minipage}
\hfill
    \begin{minipage}[t]{0.50\linewidth}
    \includegraphics[width=1\linewidth]{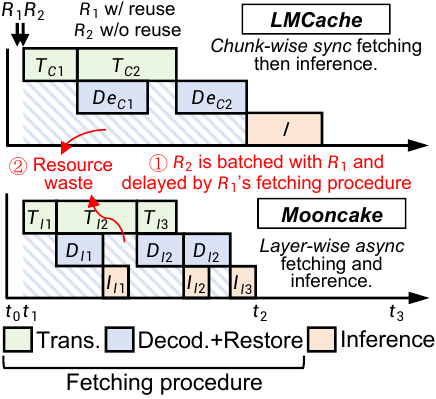}
    \caption{SOTA KV management systems deliver inefficient KV fetching on regular (non-RDMA) networks.}
      \label{fig:NVDECUTL}
\end{minipage}
\vspace{-1em}
\end{figure}


\textbf{C1: Difficult accuracy-compression tradeoff.\label{sec:InefficientCompress}}
To minimize transmission latency, the system must maximize the compression ratio while guaranteeing the inference accuracy. 
However, naively encoding KV tensors with video codecs results in a suboptimal tradeoff. 
To demonstrate this, we compare three SOTA methods CacheGen~\cite{liu2024cachegen}, ShadowServe~\cite{xiang2025shadowserve}, and llm.265~\cite{xu2025llm265}, against three encoding configurations applied on KV tensors\footnote{All three configurations first pad (\texttt{nn.ZeroPad2d}) KV cache in $[N,32,32,128]$ shape, then reshape (\texttt{torch.reshape}) it to $[N,256,176,3]$, \ie, $N$ of $[256,176,3]$ video frames, lastly encode them into video in H.265 via NVENC.} as illustrated in Fig.\ref{fig:encode_flow}: (1) \textit{Default} (standard NVENC settings), (2) \textit{QP0} (set quantization parameter as zero), and (3) \textit{Lossless} (bypassing the lossy steps). 

As shown in Fig.\ref{fig:AccComp}, all these methods fail to balance accuracy and compression ratio.
({\romannumeral 1}) \textit{Lossy step destroys accuracy.} 
Default, QP0, and llm.265 achieve high compression ratios with DCT and quantization, but suffer from accuracy drops.
This is because of the different sensitivity of high-frequency information between human eyes and LLM inference.
The high-frequency values imperceptible to human eyes and smoothed out by DCT and quantization typically correspond to activation outliers in LLMs, which act as critical attention sinks or salient features that contain essential information for accurate inference~\cite{dettmers2022llmint8, xiao2024streamingllm}.
(\romannumeral 2) \textit{The Lossless yields low compression ratios.}
While preserving accuracy, Lossless configuration only delivers a comparable compression ratio to CacheGen and ShadowServe. 
This implies it fails to exploit the inter- and intra-frame compression gain\footnote{Video encoder exploits spatial and temporal redundancy through intra- and inter-frame prediction, eliminating them when pixels can be accurately predicted from neighboring pixels or reference frames, storing only the residual between the actual pixels and the prediction.}, degenerating into a simple entropy coder.
We attribute this failure to the naive mapping from KV tensors to video frames, which disrupts the inherent spatial-temporal redundancies (more in \S\ref{sec:KVCompress}).
Crucially, this result misled llm.265 to conclude that KV tensors lack temporal similarity and incorrectly discard the inter-frame prediction step as shown in Fig.\ref{fig:encode_flow}.
\textbf{C2: Inefficient remote KV fetching.}\label{sec:IllconsideredRegionEnhancementb}
While compact compression enables fast transmission of video chunks, efficient KV fetching remains a challenge. 
Naively integrating video codecs into existing remote KV reuse systems~\cite{qin2025mooncake, lmcache} fails to meet this goal as three reasons. 
({\romannumeral 1}) \textit{Scheduling interference (HOL Blocking)}. 
The schedulers of current systems naively orchestrate the fetching requests with normal ones (\eg, $R_1$ with $R_2$ in Fig.\ref{fig:NVDECUTL}) in one batch.
This fetching-agnostic policy severely blocks non-reuse requests from inference, increasing the TTFT.
({\romannumeral 2}) \textit{Pipeline stalls and hardware underutilization}.
As shown in Fig.\ref{fig:NVDECUTL}, LMCache's inference-blocking fetch policy leaves GPU compute resources idle until receive all KV cache. 
Mooncake boosts the efficiency by a layer-wise fetching-inference pipeline.
But, it still wastes resources because it lacks mechanisms to handle the networking jitter (\eg, the longer T$_{l2}$ in Fig.\ref{fig:NVDECUTL} when bandwidth drops) and underutilizes decoding hardware (\ie, <20\% NVDEC utilization under Fig.\ref{fig:NVDECUTL} setting).
({\romannumeral 3}) \textit{KV restoring memory contention}.
LMCache and Mooncake restore original KV tensors from compressed form at chunk granularity.
This coarse-grained design causes sudden memory spikes (\eg, 2GB per chunk, details in \S\ref{sec:decompression}), which contend with the inference memory access, worsening the TTFT.

\section{\name Design}
\name is an efficient remote KV cache reuse solution that enables KV coding with GPU-native codec hardware by addressing the above challenges.
We first provide an overview of \name, then describe the core techniques it leverages.

\subsection{System Overview}

\begin{figure}[t]
\centering
\includegraphics[width=1\linewidth]{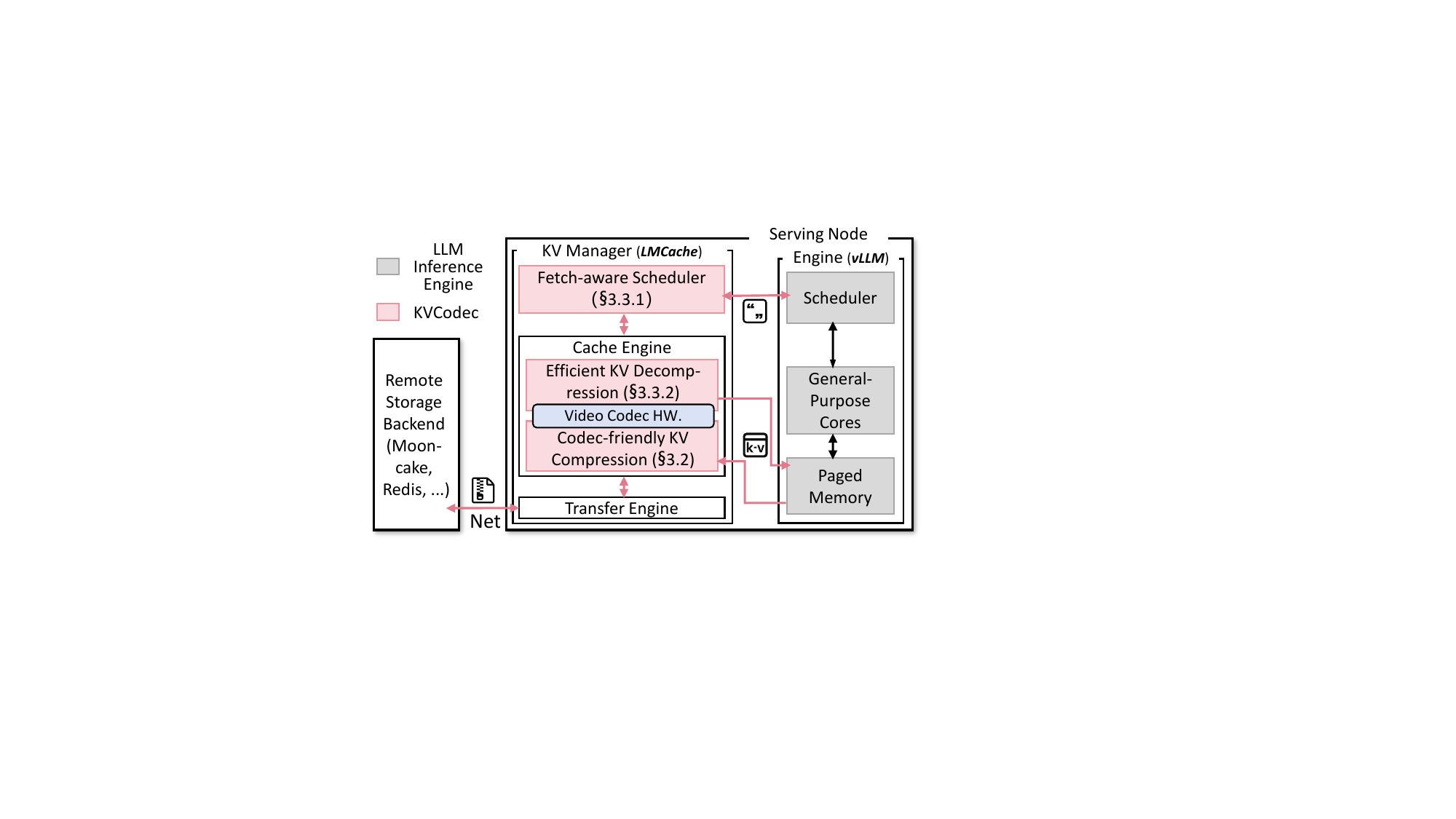}
    \caption{\name overview.}
    \label{fig:overview}
    \vspace{-1.5em}
\end{figure}

\name integrates three modules into the original KV cache manager, as shown in Fig.\ref{fig:overview}.
Fetching-aware scheduler (\S\ref{sec:Scheduler}) collaborates with the LLM inference engine, distinguishes the requests that need remote KV caches, and instructs the cache engine for KV fetching.
KV decompression (\S\ref{sec:adaptivefetch}), triggered by the scheduler, fetches video chunks from remote storage nodes, rapidly decodes them to video frames, restores the frames to the original KV caches, and writes them into the paged memory of the LLM engine for inference.
KV compression ingests the KV tensors from the inference engine, reshapes them to a codec-friendly tensor layout (\S\ref{sec:KVCompress}), and encodes them into videos.  
These KV videos are delivered to storage servers or stored in the local KV cache pool and registered as reusable.
KV compression and decompression execute in the background, transparent to inference engines.
\name follows the same setup as prior studies~\cite{liu2024cachegen, xiang2025shadowserve, qin2025mooncake}, where KV caches are chunked and encoded offline.

\subsection{Codec-friendly KV Compression}\label{sec:KVCompress}
\begin{figure*}[t]
\centering
\setlength{\abovecaptionskip}{5pt}
  \begin{minipage}[t]{0.22\linewidth}
    \includegraphics[width=1\linewidth]{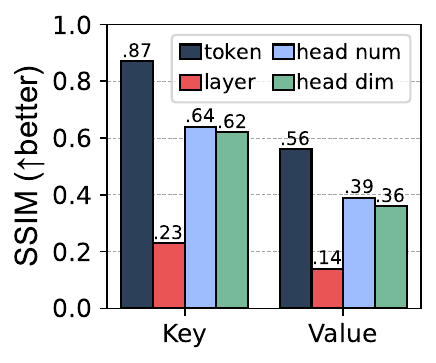}
    \caption{Similarity analysis of slicing KV cache along different dimensions.
    }
      \label{fig:ssim}
\end{minipage}
\hfill
    \begin{minipage}[t]{0.21\linewidth}
    \raisebox{0.1cm}{\includegraphics[width=1\linewidth]{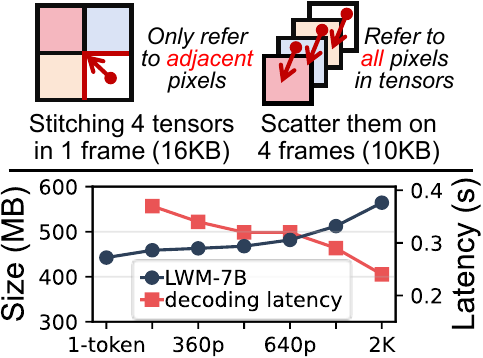}}
    \caption{Tensor encoding of diverse placement (top) and resolutions (bottom).
    }
      \label{fig:tensorencoding}
\end{minipage}
\hfill
  \begin{minipage}[t]{0.55\linewidth}
  \centering
    \includegraphics[width=1\linewidth]{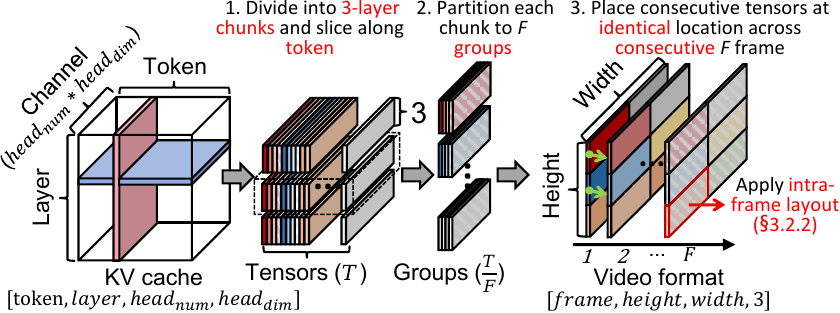}
    \caption{Inter-frame layout splits the KV cache into tensors and distributes them onto video frames to create the most temporal redundancy.
    The layout of each tensor is in \S\ref{sec:intralayout}.
     }
      \label{fig:InterLayout}
\end{minipage}
\vspace{-1em}
\end{figure*}

To minimize the transmission delay, KV caches must be compressed in a highly compact format.
The typical compression gain of video encoding comes from the lossless redundant image content elimination and the lossy numerical approximation.
To ensure LLM generation quality, \name skips lossy steps while aiming to fully leverage lossless redundancy elimination.
To this goal, it necessitates an appropriate mapping to map the KV cache with shape {\small $[token, layer, head_{num}, head_{dim}]$}, as shown the left in Fig.\ref{fig:InterLayout}, to video format with shape {\small $[frame, height, width, 3]$} as the right in Fig.\ref{fig:InterLayout}, allowing it to be encoded in a minimum size.

However, the simple mappings deliver very low compression ratios.
For example, map KV cache to {\small $layer/3$} number of video frames with shape {\small $[token, channel, 3]$} (where {\small $channel=head_{num}\times head_{dim}$}) like llm.265, \ie, slice the KV cache in Fig.\ref{fig:InterLayout} horizontally and serve every three continuous layers as one frame, and encodes them to H.265 bitstream, yielding only 58\% of ours compression ratio.
Similarly, slicing the KV cache along the token dimension, \ie, vertically in Fig.\ref{fig:InterLayout}, as in CacheGen, delivers only 42\% of ours.

To maximize the compression ratio, we propose a \textit{codec-friendly tensor layout} comprising two stages.
In inter-frame layout (\S\ref{sec:InterLayout}), it determines how to split the KV cache and where to place the resulting tensors on each frame to expose the most redundancy that video encoding can leverage.
In intra-frame layout (\S\ref{sec:intralayout}), it searches for the optimal tensor shape and inner-tensor element permutation to achieve the maximum compression ratio.

\subsubsection{Inter-frame layout.}\label{sec:InterLayout}
To guide the inter-frame layout, we first analyze the characteristics of the KV cache and video encoding, deriving three observations.


({\romannumeral 1}) \textit{Slicing KV cache along token dimension yields the highest image similarity.}
To identify the axis providing the most redundancy, we slice the KV cache along each dimension, treat sequential slices as consecutive frames, and evaluate their visual similarity with SSIM and PSNR\footnote{Structural Similarity Index Measure (SSIM) and Peak Signal to Noise Ratio (PSNR) are the most widespread metrics to measure similarity of two images. The evaluation result of PSNR and visualization is detailed in Appx~\ref{sec:similarityadditional}.}.
As shown in Fig.\ref{fig:ssim}, token dimension yields the highest similarity scores.
We attribute this to the architectural properties of LLM.
Modern LLMs' causal self-attention injects information from preceding tokens into the subsequent ones, which makies them blend with each other; and the similar positional encoding of neighboring tokens further brings them closer. 

({\romannumeral 2}) \textit{Placing a set of tensors over multiple frames delivers more compression gain than on one single frame.}
As shown in Fig.\ref{fig:tensorencoding}(top), given four consecutive tensors sliced from the KV cache, serving them as four consecutive frames and encoding them into a video yields a $1.6\times$ compression gain compared to stitching them in
a single frame.
This benefit stems from the differential compression nature of video encoding.  
When stitching into a single frame, each pixel block (tensor) can only refer to boundary pixels from its adjacent left and upper blocks (\eg, the red lines in Fig.\ref{fig:tensorencoding}(top)), which wastes many referable opportunities within the blocks.
In contrast, when serving the tensors as four consecutive frames, the pixel block can refer to all pixels in its predecessors, resulting in a higher compression ratio.


({\romannumeral 3}) \textit{Compression ratio is sensitive to the video resolutions.}
It seems we can easily serve each tensor as a video frame to minimize video size and transmission delay; however, such a video cannot be decoded (144P is the smallest feasible resolution for NVDEC).
As shown in Fig.\ref{fig:tensorencoding}(bottom), while the video size increases with higher resolution as more tensors are stitched into a single frame, the decoding efficiency also benefits. 
Given this complex relationship, video resolution should be carefully set.



With these observations in mind, inter-frame layout obeys two principles.
(1) Slice KV cache along the token dimension and place the token-adjacent tensors onto continuous frames (observations ({\romannumeral 1}) and ({\romannumeral 2})), constructing the maximum temporal redundancy.
(2) Encode videos in multiple-resolution versions. 
Runtime can adaptively select the sweet-spot resolution that minimizes TTFT (observation ({\romannumeral 3})), balancing transmission and decoding time (\S\ref{sec:adaptivefetch}). 

Following these two principles, our inter-frame layout consists of three steps as illustrated in Fig.\ref{fig:InterLayout}.
1) It first divides the KV cache into three-layer chunks and slices them along the token dimension; each chunk contains $T$ tensors, each with shape $[1, 3, channel]$.
2) $T$ tensors in each chunk, \eg, the dashed box in Fig.\ref{fig:InterLayout}, are sequentially partitioned into $\frac{T}{F}$ groups, each containing $F$ tensors.
3) Adjacent tensors in each group are placed at identical positions across consecutive $K$ frames, and the best tensor layout is determined with intra-frame layout.
Such an inter-frame layout maintains the spatial alignment of consecutive token tensors, enabling the encoder to predict each tensor by referencing its temporal predecessor (green arrows in Fig.\ref{fig:InterLayout}), thereby maximizing temporal redundancy.
Moreover, the three layers (lowest similarity in Fig.\ref {fig:ssim}) are mapped to independently coded color channels (YUV/RGB). 

\subsubsection{Intra-frame layout.}\label{sec:intralayout}
\begin{figure}[t]
\centering
\setlength{\abovecaptionskip}{5pt}
    \includegraphics[width=1\linewidth]{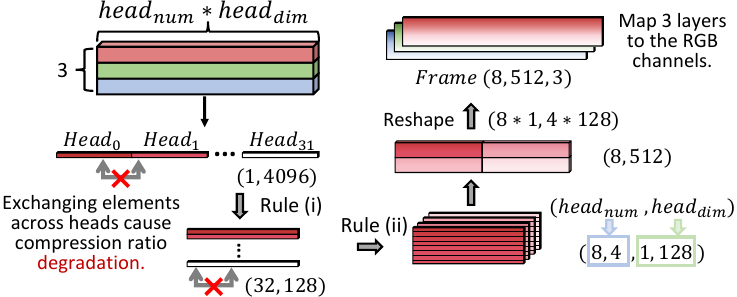}
    \caption{Intra-frame layout searches for the optimal mapping for the tensor providing the max compression ratio.
    }
      \label{fig:IntraLayout}
\vspace{-1em}
\end{figure}

After slicing the KV cache into three-layer token tensors and placing them in the appropriate locations,  
our next goal is to determine the mapping of remaining {\small $head_{num}$ (H)} and {\small $head_{dim}$ (D)} dimensions.
Keeping the original tensor layout without adjustment, \eg, the {\small $[1, 3, head_{num}\times head_{dim}]$} as shown the upper left in Fig.\ref{fig:IntraLayout}, can only deliver an 8.7$\times$ compression ratio, much smaller than our ultimate 11.9$\times$.
Therefore, we must search for the best intra-frame layout to optimize the compression ratio.

We formulate the inter-frame layout as a joint optimization problem that couples geometric tiling (tensor reshaping) with inner-element permutation.
It creates a {\small$O(\log N \times N!)$} of solution space, {\small$N = H \times D$}, where brute-force searching is intractable.
As exemplified in Fig.\ref{fig:IntraLayout}, given a $(1,4096)$ vector of one-layer tensor, there exists 13 geometric tiling 
times $4096!$ permutation solutions.
Enumerating such a huge number of possibilities and evaluating the video size of each is impossible.

To reduce the search space, we follow three rules deriving from the characteristics of LLM architecture.

Rule ({\romannumeral 1}): Do not exchange the elements from different attention heads.
In LLMs, each attention head captures a unique semantic feature.
Exchanging elements between one head and another disrupts this semantic information.
It breaks the similarity among continuous tokens and inter-frame prediction, leading to a large residual.
Our experimental results demonstrate that exchanging 50\% of elements across 32 attention heads results in a 2.4$\times$ degradation in compression ratio.
According to this, we isolate the element permutation across attention heads and separately consider {\small $head_{num}$} and {\small $head_{dim}$}, as $(1,4096)$ to $(32,128)$ shown in Fig.\ref{fig:IntraLayout}.
It eliminates enumerating all 4096! permutation, reducing the entire searching space from {\small$O(\log N \times N!)$} to {\small$O((\log H \times H!)\times(\log D \times D!))$}.

Rule ({\romannumeral 2}): Keep the order of elements within the attention head. 
As the attention head serves as the fundamental semantic unit of LLMs, elements within a head jointly represent a specific feature.
Disarranging the elements of the attention head disrupts its structured feature correlations, hindering intra-frame prediction. 
In our experiments, exchanging 50\% of elements in the attention head increases the 17\% size of intra-predicted frames.
For this reason, we preserve the inner-head element order, which omits the elements permutation of each attention head and reduces the searching space from {\small$O((\log H \times H!)\times(\log D \times D!))$} to {\small$O((\log H \times H!)\times\log D$}.

Rule ({\romannumeral 3}): Arrange the order of attention heads as initial.
Since distinct heads extract independent semantic features without an explicit order, their relative positions have a negligible impact on compression.
Our empirical studies demonstrate that random head orders yield only <0.3\% of size variation.
Accordingly, we do not permute the attention heads but solely search the geometric tiling of {\small $head_{num}$}, which ultimately reduces the search space to very limited {\small$O(\log H\times\log D)$}.
As shown in Fig.\ref{fig:IntraLayout}, it only requires evaluating the compression ratio of $\log 32\times\log128=35$ possibilities.



\begin{figure}[t]
\centering
\setlength{\abovecaptionskip}{5pt}
\includegraphics[width=0.86\linewidth]{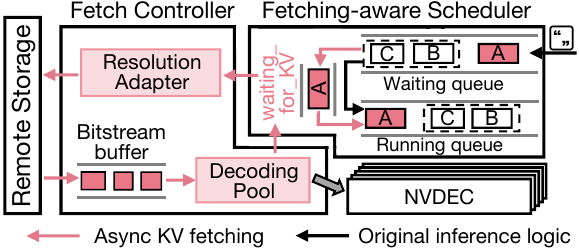}
    \caption{Control flow of KV fetching. Fetching-aware scheduler discriminates the requests. Fetch controller isolates the KV fetching procedure.}
    \label{fig:controlflow}
    \vspace{-1em}
\end{figure}

Based on the principles above, the optimal layout converges to only a few dozen options. 
Since all these principles depend solely on the model architecture and video encoding, namely being input-agnostic (more in \S\ref{sec:ComponentwiseAnalysis}), we can search for the best layout offline. 
Spending only 1.5 hours, we obtain the best layouts $(8,512)$, $(8,128)$, and $(16,64)$ for all three models used in this paper,  LWM-7B, Yi-34B, and Llama-70B,
As exemplified in Fig.\ref{fig:IntraLayout}, the {\small $head_{num}$} and {\small $head_{dim}$} of LWM-7B are reshaped to $(8, 4)$ and $(1, 128)$, and then further reshape to $(8, 512)$.
Lastly, three $(8, 512)$ matrices from each layer are batched into a $(8, 512, 3)$ tensor. 

\subsection{Efficient Remote KV Fetching}\label{sec:Fetching}
To minimize TTFT, we propse three key techniques, enabling \name efficient remote KV cache fetching. 







\subsubsection{Fetching-aware scheduler.}\label{sec:Scheduler}
Existing remote KV reuse systems batch requests without discrimination, causing requests with remote KV fetch to block the inference of non-reuse requests (\S\ref{sec:challenges}).
To tackle this, KV fetching must be isolated from the inference engine's main execution flow.
We propose a \textit{fetching-aware scheduler} that enables asynchronous KV fetching, avoiding blocking non-reuse requests.

As shown in Fig.\ref{fig:controlflow}, our scheduler imports a dedicated queue, \texttt{waiting\_for\_KV}, to manage fetching requests.
It is located outside the LLM inference engine and collaborates with the original queues.
In each iteration, the scheduler distinguishes the fetching requests, moves the eligible ones (
\eg, request A) from the \texttt{waiting} queue into the \texttt{waiting\_for\_KV} queue, and notifies the fetch controller to start fetching their KV caches in the background.
The non-reuse requests (\eg, B and C) still follow the original logic of LLM inference engine, entering the \texttt{running} queue for immediate inference.
Once fetching completes, the fetch controller asks the scheduler to dequeue request A from \texttt{waiting\_for\_KV} queue to \texttt{running} for immediate execution in the next iteration.
Benefiting from this design, the inference engine remains unaffected by KV fetching, maintaining its non-reuse or user-defined scheduling policy and computation.



\subsubsection{Efficient KV Decompression.}\label{sec:decompression}
\begin{figure}[t]
\centering
\setlength{\abovecaptionskip}{5pt}
\includegraphics[width=0.48\textwidth]{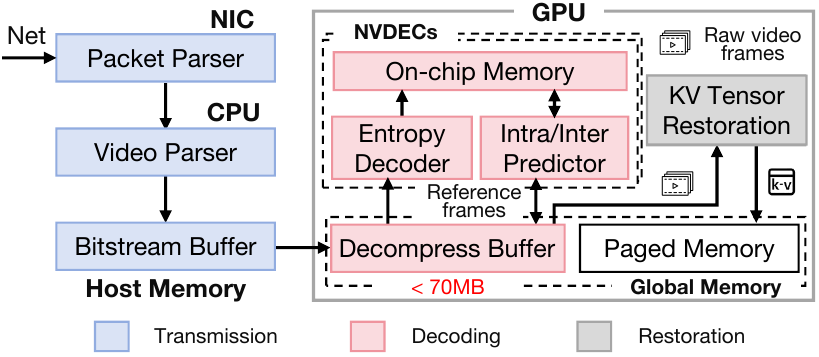}
    \caption{Dataflow of KV fetching, consisting of KV video transmission, KV video decoding, and KV tensor restoration.}
    \label{fig:dataflow}
    \vspace{-1em}
\end{figure}

Remote KV fetching 
consists of three phases, involving multiple hardware devices. 
During the transmission phase, the resolution adapter in Fig.\ref{fig:controlflow} specifies the KV video chunk, fetches it from remote storage, and stores it in the bitstream buffer in host memory (see Fig.\ref{fig:dataflow}). 
During the decoding phase, the decoding pool, each instance combined with an NVDEC unit, ingests video chunks and decodes them to frames.
Lastly, the restoration module restores decoded frames to original KV tensors and writes them to the paged memory of LLM inference engines.

\begin{figure}[t]
\centering
\setlength{\abovecaptionskip}{5pt}
\includegraphics[width=1\linewidth]{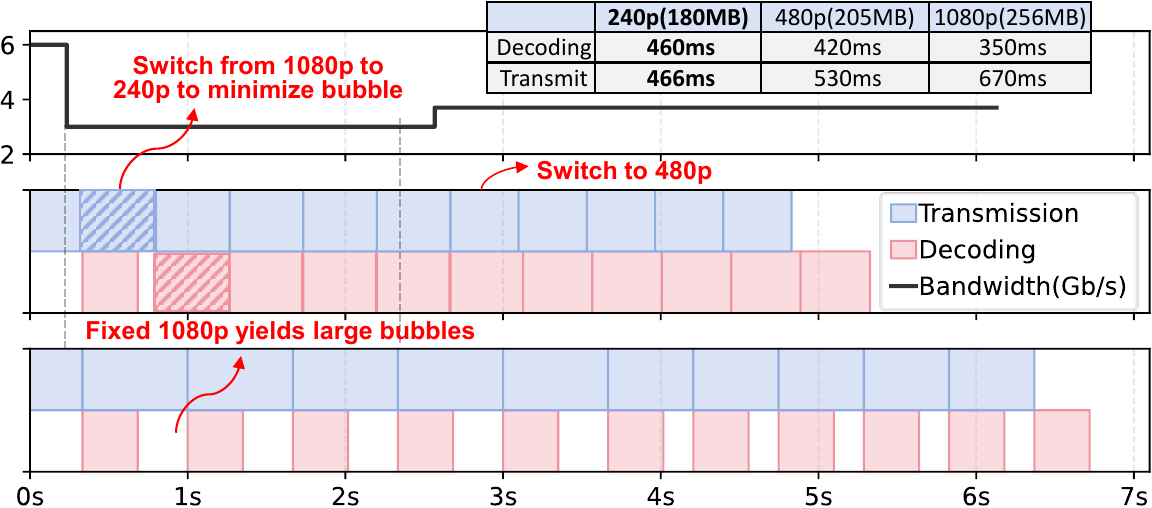}
    \caption{Adaptive resolution KV fetching significantly reduces the pipelining bubbles between transmission and decoding under dynamic bandwidth. }
    \label{fig:adaptiveresolution}
    \vspace{-1em}
\end{figure}

\textbf{Adaptive resolution KV fetching}.\label{sec:adaptivefetch} 
To minimize TTFT, we employ a \textit{pipelining} mechanism to overlap transmission and decoding (restoration naturally overlaps with decoding, elaborated shortly).
However, this is non-trivial for two reasons.
First, the transmission and decoding efficiency exhibit opposite characteristics.
While transmission favors low-resolution videos (\eg, 240P in observation ({\romannumeral 3}) of \S\ref{sec:InterLayout}), it fails to saturate the block-parallel (64$\times$64 pixels) decoding units~\cite{sullivan2012H265}, incurring $1.3\times$ higher decoding latency than 1080p as the table shown in Fig.~\ref{fig:adaptiveresolution}.
Second, the entire KV fetch may take dozens of seconds, during which network conditions are very likely to fluctuate.
Static video chunk sizes suffer from dynamic bandwidth, leading to pipeline bubbles.




To remove these bubbles, we propose a \textit{bandwidth-aware resolution adaptation} mechanism.
Unlike CacheGen, which pursues transmission efficiency through aggressive quantization at the cost of accuracy, our approach adjusts the tensor-to-frame layout (\ie, resolution) and so the video chunk size, adapting to dynamic bandwidth without accuracy loss.
As modern GPUs are equipped with abundant NVDECs (\eg, 5 NVDECs per A100), we abstract them into a decoding pool, as shown in Fig. \ref{fig:controlflow}, to simultaneously decode multiple video chunks. 
Once a decoding instance is idle, one chunk is dequeued from the bitstream buffer for immediate decoding.


Our adaptation mechanism uses a profile-based table-lookup method to select the optimal resolution for the video chunk to be fetched.
When fetching a video chunk, the resolution adapter predicts the network bandwidth (for simplicity, calculated from the last chunk's transmission delay), traverses the transmission latency for all resolutions, estimates decoding delay by table look up, and then selects the optimal resolution that delivers the minimum bubble (Appx.\ref{sec:pseudocode} details pseudo code and complete lookup tables).
To make it concrete, Fig.\ref{fig:adaptiveresolution} illustrates an example. 
Before fetching the 2nd chunk, the bandwidth degrades from 6 to 3 Gbps. 
Resolution adapter estimates the transmission delay for all resolutions, looks up their decoding times (as shown in the table), and selects the optimal 240p that delivers the smallest 6ms bubble. 
Four chunks later, bandwidth increases to 4 Gbps, so the adapter switches to 480p.
Compared to the fixed 1080p approach, our adaptive-resolution method eliminates most bubbles, thereby saving 21\% in time cost.

\textbf{Frame-wise KV tensor restoration.}\label{sec:restoration}
During the decoding phase, as illustrated in Fig.\ref{fig:dataflow}, the entropy decoder decomposes the bitstream into intra- and inter-frame parts and stores them into on-chip memory; then, the intra- and inter-frame predictor reads them out to reconstruct raw video frames.
Due to the limited on-chip memory capacity, inter-frame prediction typically keeps the reference frames in GPU global memory, competing with the LLM inference engine.

To achieve interference-free decoding, we propose \textit{frame-wise KV restoration}, which minimizes reliance on global memory via two strategies.
(1) With careful tensor layout in KV compression, the number of reference frames is limited to less than four frames.
This significantly reduces memory overhead to less than 20MB, even at 2K resolution.
(2) Unlike CacheGen and ShadowServe that restore original KV tensors in chunks (1.5K tokens per chunk), 
our approach executes frame-wise.
Once a raw video frame is decoded, it is immediately mapped back to the original KV tensors and filled into the preallocated slots in the paged memory.
Some local cache policies~\cite{gao2025fast, yu2025stateful} store the original KV tensors in host memory or disks.
We follow the CacheGen and Shadowserve, filling them into paged memory directly.

We implement this as a callback function, \texttt{On\_frame\_probe}.
This delivers two-fold advantages.
On the one hand, the callback function can be plugged into the codec, naturally pipelining with decoding phase.
On the other hand, such a frame-wise restoration brings remarkable memory reduction to the decompress buffer, from the prior chunk-wise 1.5-2GB to the frame-wise 50MB.
Consequently, combining the memory cost of reference frames, the overall decompress buffer is less than 70MB.
Notably, although \texttt{On\_frame\_probe} executes on CUDA cores, it is super lightweight with only a few reshape operations. 
The frame-to-tensor mapping 
are on-hand encoded in the bitstreams during KV compression. 

\vspace{-0.5em}
\section{Implementation}
We implement \name as a pluggable backend of LMCache(v0.3.7)~\cite{lmcache}, and use vLLM(v0.10.2)~\cite{kwon2023efficient} as the LLM inference engine.
The codebase primarily uses Python (v3.12), and the efficient KV fetcher is partially implemented in C++ and CUDA, totaling $\sim$5K LOC.
(1) The KV compression is implemented with FFmpeg~\cite{ffmpeg} API interface with NVENC, and it falls back to CPU encoding when lacking NVENCs. 
The encoding follows H.265 with \texttt{lossless=1} to skip lossy steps. 
As with CacheGen, the KV cache is quantized to integers and encoded into video chunks (each containing 10K tokens across three layers) in multiple resolutions, before being registered as reusable.
(2) For efficient KV fetching, the fetching-aware scheduler runs on a standalone thread integrated into LMCache, synchronizing with the vLLM scheduler via thread events. 
Adaptive resolution adaptation is implemented simply in Python, but for concurrent video decoding, all NVDECs are abstracted into a resource pool, with each instance pinned to an NVDEC for a decoding pipeline built on GStreamer~\cite{GStreamer} via \texttt{nvv4l2decoder}.
The decoding pipeline is implemented in C++ and invoked by Pybind11 with \texttt{py::gil\_scoped\_release} to bypass the Python GIL, ensuring that decoding does not block the vLLM main loop, and frame-wise KV tensor restoration is implemented as a 
\texttt{On\_frame\_probe} callback plugged in it, with a customized \texttt{Sparse\_frame\_KV\_transfer} operator fast write KV tensors into paged memory.
When multiple fetching requests are routed to a single serving node, \name applies a FCFS policy if a request consumes the entire bandwidth or GPU memory; otherwise, it batches concurrent requests and partitions bandwidth evenly as CacheGen did for their KV fetching.

\textbf{Compatibility.}
Since the seamless integration between LMCache and vLLM, \name maintains full compatibility with vLLM's native features.
Notably, unlike LMCache's blocking inference for fetching, \name follows Mooncake's layer-wise fetching-inference pipelining.
It immediately enqueues fetching requests into vLLM's \texttt{running} queue, provided the remaining layers' fetching time can be hidden by inference.
Thanks to the inherent chunked prefill from vLLM, the inference time cost can be precisely estimated, hence eliminating the pipelining bubbles of this layer-wise design.
For more details, please see Appdx.\ref{sec:layerwisepipeline}.




\section{Evaluation}\label{sec:evalutaion}
We evaluate \name on three heterogeneous GPU clusters with real-world request traces. 
The key takeaways are:

\scalebox{0.8}{$\bullet$} \name reduces up to 3.51$\times$ TTFT for fetching requests and 77\% TTFT and 35.4\% TPOT for non-reuse requests. The performance gains are consistent across diverse models, GPUs, and network bandwidths. (\S\ref{sec:E2EPerf})

\scalebox{0.8}{$\bullet$} Codec-friendly KV compression and adaptive-resolution KV fetching bring remarkable bandwidth and TTFT savings, and frame-wise KV tensor restoration guarantees the interference-free inference. (\S\ref{sec:ComponentwiseAnalysis})



\subsection{Experimental Setup}\label{sec:ExperimentalSetup}
\textbf{Models.} We evaluate \name on three models of different sizes and context capabilities: 
LWM-7B~\cite{lwm7b} with 1M, Yi-34B~\cite{yi34b} with 200K, and Llama3-70B~\cite{llama70b} with 128K.

\textbf{Datasets.} We evaluate \name on three long-context benchmarks to comprehensively assess performance across various tasks and context lengths.
(1) \textit{L-Eval}~\cite{an2024eval} contains 20 closed-ended QA tasks, including 508 long documents with lengths ranging from 3-200K tokens.
(2) \textit{LV-Eval}~\cite{yuan2024lv} further increases difficulty and mitigates knowledge leakage, consists of single-hop and multi-hop QA of distraction documents, fact confusion, and keyword replacement with lengths ranging from 16-256K. 
(3) \textit{LongBench-V2}~\cite{bai2025longbench} is the most systematic evaluation that employs a multiple-choice format to ensure objectivity across single- and multi-document QA, long-form conversations, coding, and structured data tasks, with 13-167K context lengths.


\begin{figure*}[t]
    \centering
    \setlength{\abovecaptionskip}{5pt}
    \begin{minipage}[b]{1\linewidth}
    \centering
    \includegraphics[width=0.8\linewidth]{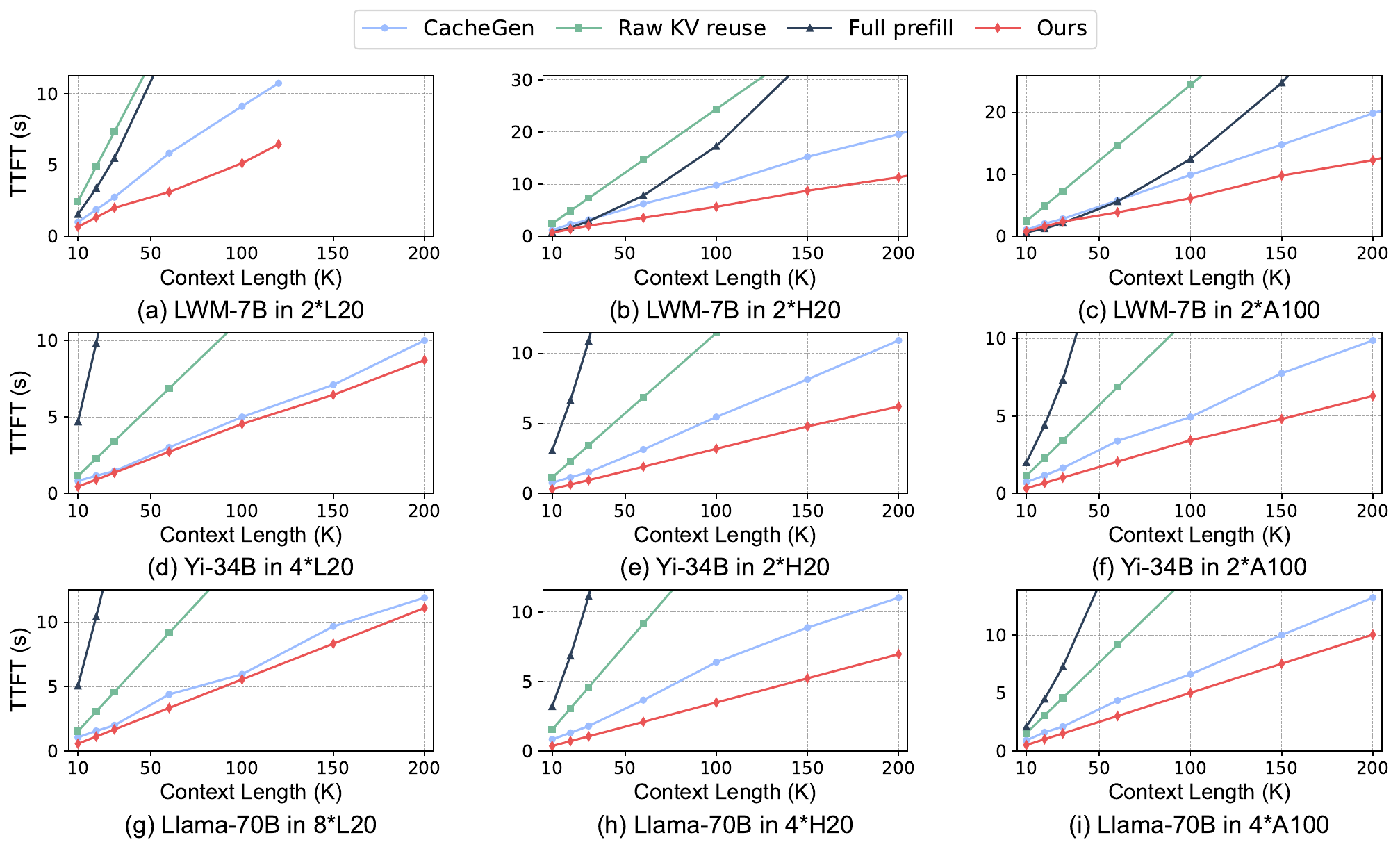}
    \end{minipage}
    \caption{TTFT of the request with remote KV reuse across different context lengths over various devices and models.}
    \label{fig:ttft_performance}
\end{figure*}

\textbf{Metrics.} We compare \name's \textit{accuracy}, \textit{latency}, and \textit{compression ratio} with baselines.
Accuracy follows the standard metric of each dataset.
For L-Eval and LongBench, it is the percentage of generated answers that correctly match the ground truth; for LV-Eval, it is the F1 score, harmonic mean of precision and recall, between the generated answer and the ground-truth answer for the QA task.
Latency and compression ratio still follow \S\ref{sec:challenges}.


\textbf{Baselines.} To show the superiority brought by \name, we compare it with: \textit{Full prefill} executes the standard LLM inference without any KV cache reuse; 
\textit{Raw KV reuse} pulls the raw KV cache remotely and computes their cross attention with query prompts;
and three Compressed KV reuse methods, \textit{CacheGen}, \textit{ShadowServe}, and \textit{llm.265}.



\textbf{Test platform.} 
We test on three GPUs representing different market positions, each with a different number of cards, supporting different model sizes.
For high-end NVIDIA A100 (80GB) with 5 NVDECs and mid-range NVIDIA H20 (96GB) with 7 NVDECs, LWM, Yi, and Llama3 leverage 2, 2, and 4 cards, respectively.
For low-end NVIDIA L20 (48GB) with 3 NVDECs, three models utilize 2, 4, and 8 cards.

\subsection{End-to-End Performance}\label{sec:E2EPerf}
This section reports the E2E performance of \name on various devices.
If not specified, all results are tested on H20 with the Yi-34B model over a 16Gbps bandwidth network, which is typically offered by regular cloud platforms~\cite{xiang2025shadowserve}.


\begin{figure}[t]
    \centering
    \setlength{\abovecaptionskip}{0pt}
    \subfigure[TTFT] {
     \label{fig:4090}
    \begin{minipage}[t]{0.45\linewidth}
    \centering
    \includegraphics[width=1\linewidth]{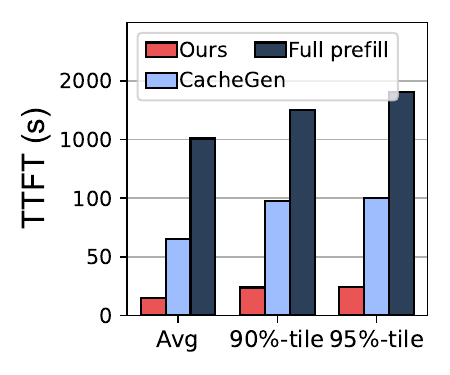}
    \end{minipage}
    }
    \hfill
    \subfigure[TPOT] {
    \begin{minipage}[t]{0.45\linewidth}
    \centering
    \includegraphics[width=1\linewidth]{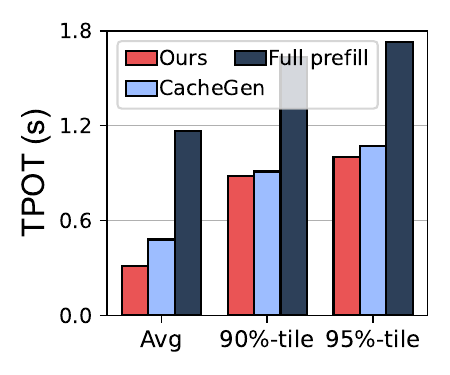}
    \end{minipage}
    }
    \caption{\name's KV fetching provides remarkable TTFT gain to non-reuse requests. \name also offers TPOT benefit under PD-aggregated environments.}
    \label{fig:multi_req_perf}
    \vspace{-1em}
\end{figure}

\noindent
\textbf{TTFT saving of requests requiring KV fetching.} 
As illustrated in Fig. \ref{fig:ttft_performance}, \name achieves the lowest TTFT for the fetching request across various hardware and models with varying context lengths.
It outperforms Full prefill, Raw KV reuse, and CacheGen by $13.63\times$, $3.51\times$, and $1.52\times$, on average.
Full prefill is extremely inappropriate for long-context LLM serving; the superlinear computational complexity with context lengths causes unacceptable TTFT.
Although raw KV reuse and CacheGen reuse remote KV caches to mitigate this heavy computation, their TTFTs are bottlenecked by transmission over regular cloud bandwidth because they lacks efficient KV compression.
In contrast, \name achieves highly compact KV compression, thereby significantly reducing transmission latency. 
The most limited speedup, as shown in Fig.~\ref{fig:ttft_performance} (d) and (g), occurs at serving on the L20 GPUs. 
This phenomenon arises because L20 has only three NVDECs, which incur a queue up for decoding, and the Grouped Query Attention (GQA) used by Yi and Llama generates a relatively smaller KV cache, which reduces the benefits of our compression.
Even so, \name saves remarkable TTFT for fetching requests.

\noindent
\textbf{TTFT \& TPOT benefits for non-reuse requests.} 
Beyond TTFT reduction for fetching requests, \name keeps almost interference-free performance for non-reuse requests.
We evaluate this on a real-world request trace~\cite{qin2025mooncake} with request arrival rate at 0.2 req/s, and set $40$K tokens as the reuse threshold for \name and CacheGen, \ie, prefill requests with <$40$K context tokens and reuse remote KV for >$40$K-token requests.
All requests follow the vLLM's default First-Come-First-Served (FCFS) policy. 
As illustrated in Fig. \ref{fig:multi_req_perf}(a), \name reduces $77.1\%$ and $98\%$ of TTFT compared to CacheGen and Full prefill.
This stems from a two-fold.
For non-reuse requests orchestrated with fetching requests in a single batch, \name's fetching-aware scheduler enables KV fetching to run in the background, avoiding blocking the inference of non-reuse requests such as CacheGen and thereby reducing their TTFT.
For non-reuse requests that arrive but are still in the waiting state, all techniques in \name work together, boosting the serving efficiency of running requests, reducing the waiting time for all requests, and thus saving significant TTFT.
\name also reduce TPOT by $35.4\%$ and $40\%$ compared to CacheGen and Full prefill.
Full prefill must compute the input contexts for all requests.
Such a large amount of computing workload means vLLM must always piggyback decoding on prefilling requests, severely delaying their TPOT.
Remote KV reuse in \name and CacheGen eliminates the prefilling workload, allowing decoding to run in isolation and resulting in low TPOTs.
Moreover, compared to CacheGen's CUDA-based decompression, which competes for GPU resources, \name's codec-based method provides better performance.

\begin{figure}[t]
\centering
\setlength{\abovecaptionskip}{3pt}
\includegraphics[width=1\linewidth]{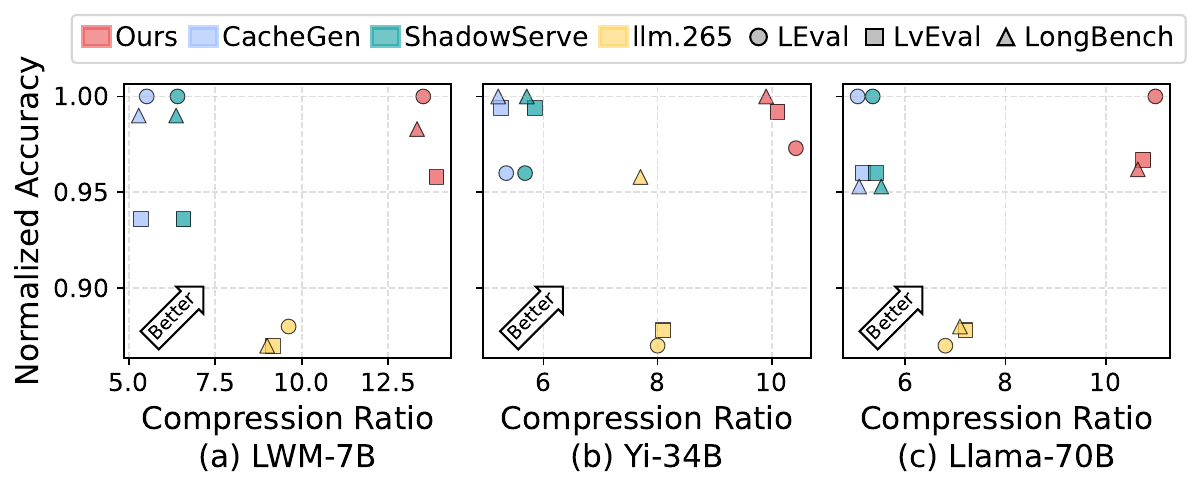}
    \caption{\name achieves the best accuracy and compression ratio tradeoff over diverse models and benchmarks.}
    \label{fig:accuracy_compression}
    \vspace{-1em}
\end{figure}

\begin{figure}[t]
\setlength{\abovecaptionskip}{3pt}
\centering
\includegraphics[width=1.\linewidth]{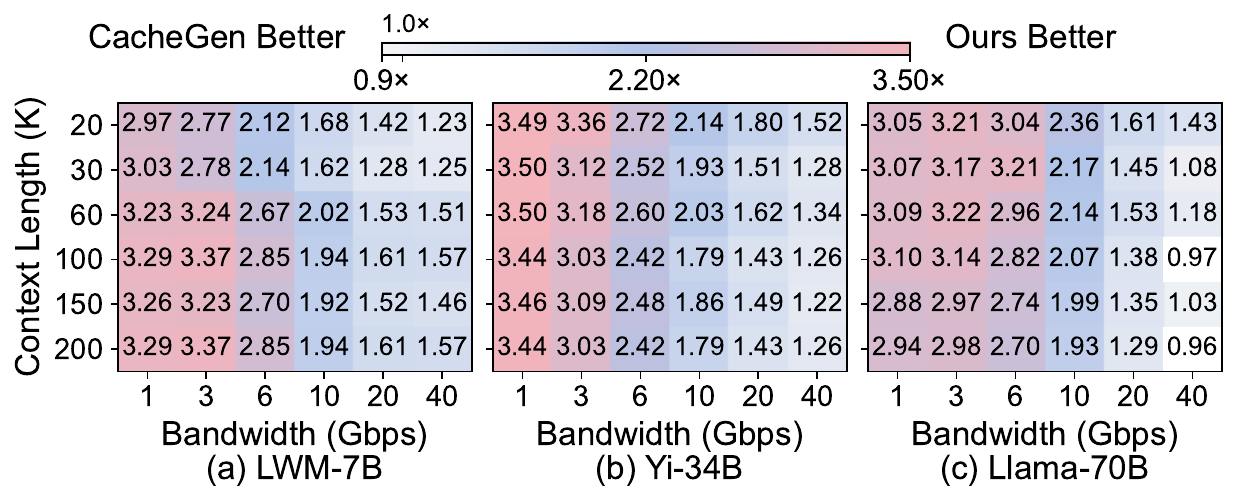}
\caption{Performance comparison between CacheGen and ours. The value is the ratio of CacheGen's TTFT $\div$ our TTFT.}
 \label{fig:CacheGenCompare}  
\hfill
\vspace{-1em}
\end{figure}

\noindent
\textbf{High compression ratio without accuracy drop.} 
As shown in Fig.\ref{fig:accuracy_compression}, \name achieves the best compression ratio across all datasets and models while maintaining lossless accuracy.
It improves the compression ratio by $2.17\times$ over CacheGen and $1.93\times$ over ShadowServe, without accuracy degradation; and against llm.265, \name delivers a $12\%$ accuracy enhancement alongside $1.41\times$ of compression ratio.
The preservation of accuracy is achieved by strictly using the H.265 lossless mode, which bypasses the lossy steps of video encoding, and the same quantization method as CacheGen and ShadowServe.
The superior compression performance stems from a codec-friendly tensor layout that appropriately maps KV tensors to pixel blocks, thereby maximizing data redundancy that can be eliminated by video coding.



\noindent
\textbf{TTFT comparison of \name v.s. CacheGen.} 
\name delivers robust performance across a wide range of bandwidth and context lengths.
We conduct a comprehensive evaluation of TTFT against CacheGen across 1-40Gbps bandwidth and 20K-200K context lengths.
Fig.~\ref{fig:CacheGenCompare} indicates that under bandwidth constraints of <40 Gbps, \name achieves an average speedup of $1.29\times$--$3.50\times$ over CacheGen.
The performance gain diminishes as the bandwidth increases.
This is because, as bandwidth increases, the arrival rate of the fetched video chunk exceeds NVDEC's decoding capacity, breaking the transmission-decoding pipeline and causing waiting (more in \S\ref{sec:ComponentwiseAnalysis}).
Conversely, CacheGen's CUDA-based decompression kernel can leverage all CUDA cores on GPUs to accelerate decompression; however, this significantly impacts the inference of non-reuse requests, as shown in Fig.\ref{fig:multi_req_perf}.
Nevertheless, in almost all scenarios, typically under low-end GPU-paired bandwidth, \name still beats CacheGen thanks to our careful design.


\begin{figure}[t]
\setlength{\abovecaptionskip}{3pt}
\centering
\begin{minipage}[b]{0.49\linewidth}
    \centering
\includegraphics[width=0.9\linewidth]{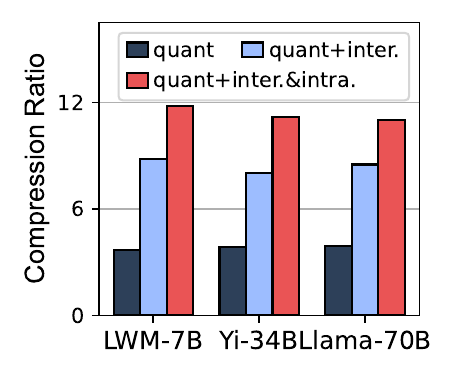}
\caption{Compression ratio breakdown on three models.}
    \label{fig:com_breakdown}
\end{minipage}
\begin{minipage}[b]{0.49\linewidth}
\centering
\includegraphics[width=0.9\linewidth]{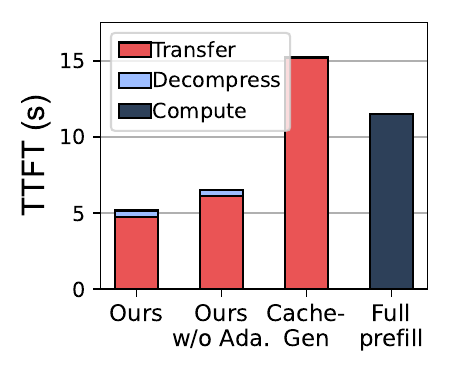}
\caption{TTFT breakdown across different baselines.}
\label{fig:ttftbreakdown}
\end{minipage}
\vspace{-1em}
\end{figure}

\subsection{Ablation Study}\label{sec:ComponentwiseAnalysis}
We evaluate the effectiveness of each component in \name by ablation studies.

\noindent
\textbf{Codec-friendly KV compression}\label{sec:ExperimentCompression} helps \name achieve 11.9$\times$ compression ratio while keeping high accuracy.
Fig.~\ref{fig:com_breakdown} breaks down the compression contribution of quantization, inter-frame layout, and intra-frame layout on all three models. 
Compared to quantization, our inter-frame layout achieves an average compression gain of 2.2$\times$, while the intra-frame layout further boosts this improvement to 2.96$\times$.
Furthermore, compared to LWM-7B and Yi-34B, the proportion of intra-frame layout in Llama-70B is the largest due to its fewest attention heads and usage of GQA.

\noindent
\textbf{Adaptive resolution KV fetching.} 
As shown in Fig.~\ref{fig:ttftbreakdown}, under the same bandwidth state as Fig.~\ref{fig:adaptiveresolution}, \name reduces the TTFT to 5.2s, representing a $20\%$ improvement over the baseline without adaptive resolution. 
This flexibility allows \name to handle network jitter without compromising accuracy, unlike CacheGen's adaptive quantization level. 
Furthermore, pipelining transmission and decoding effectively hides the decoding overhead, resulting in less than 400ms latency per video chunk.
And remote KV reuse reduces prefill computation to under 50ms.

\begin{figure}[t]
\setlength{\abovecaptionskip}{3pt}
\centering
\begin{minipage}[b]{0.49\linewidth}
\centering
\includegraphics[width=0.9\linewidth]{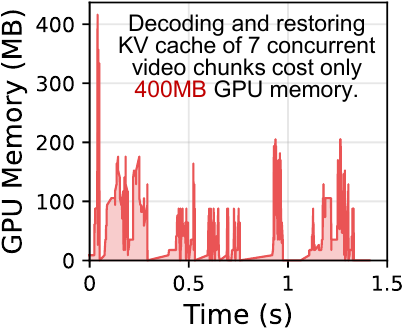}
\caption{GPU memory cost of concurrent 7 video chunks.}
\label{fig:memory_viz}
\end{minipage}
\begin{minipage}[b]{0.49\linewidth}
\centering
\includegraphics[width=0.9\linewidth]{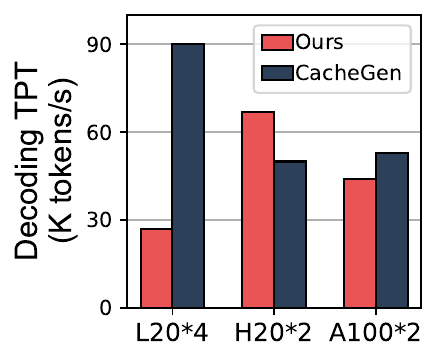}
\caption{Decoding throughput on different devices.}
 \label{fig:nvdectpt}  
\end{minipage}
\vspace{-1em}
\end{figure}

\noindent
\textbf{Frame-wise KV tensor restoration.} 
As shown in Fig.~\ref{fig:memory_viz}, concurrently decoding and restoring KV caches of 7 video chunks only occupies 400MB of peak GPU memory.
Such marginal overhead does not interfere with vLLM, which ensures memory safety in long-context scenarios.
In particular, a single KV fetching only needs NVDEC to pre-allocate 40MB for video decoding, and the restoration to reshape and dequantize with 47MB.

\noindent
\textbf{Decoding throughput}\label{sec:ExperimentNVDEC} is typically bottlenecked by available NVDEC chips.
We evaluate this with Yi-34B model on different platforms.
As shown in Fig.~\ref{fig:nvdectpt}, \name achieves 27K, 67K, and 47K tokens/s on 4 L20, 2 H20, and 2 A100 platforms.
They are only $0.3\times$, $1.34\times$, and $0.88\times$ the performance of CacheGen.
These unsatisfactory results stem from the hardware specifications: L20 has only 3 NVDECs, while H20 and A100 have 7 and 5. 
In contrast, CacheGen relies on CUDA cores, which can leverage the entire GPU computing resources.
Despite unsatisfactory, \name's overall performance is much better than CUDA-base CacheGen. 
We believe that as researchers recognize the universality of codec ASICs, next-generation GPUs will overcome these resource limitations.

\section{Limitation and discussion}\label{sec:limitation}
\textbf{Online KV compression} is essential for P-D disaggregation~\cite{zhong2024distserve}, where KV cache must be transmitted between disaggregated prefilling and decoding nodes, and for fault tolerance~\cite{fu2024serverlessllm, tumanov2016tetrisched, sun2024llumnix} or instance preemption~\cite{miao2024spotserve}, that migrates KV cache across nodes.
While \name enables compact KV compression, limited NVENC resources make the KV compression procedure insufficient to meet runtime requirements.
We believe that compressed KV reuse is promising and will support a broader range of runtime scenarios with the next-generation, powerful NVENC released.

\textbf{Preallocate GPU memory} for fetching requests.
Aligning with Mooncake and LMCache, \name treats fetching requests as first-class citizens, preallocating memory for all KV caches upfront.
Although this mechanism likely blocks the inference of non-reuse requests due to the memory budget, it ensures the timely inference of fetching requests.
One feasible improvement is to store the fetched KV cache in local storage~\cite{pan2024instinfer, xiong2024layerkv,jiang2025efficient}, making room for non-reuse request inference and swapping them to GPU memory when needed.
We leave this for future work.


\section{Related Work}\label{sec:RelatedWorks}
\textbf{LLM inference engines} today employ a bunch of techniques to boost inference performance.
Continuous batching~\cite{yu2022orca, wu2023fast} avoids ineffective zero padding by fine-grained orchestrating prefilling and decoding workloads together for higher efficiency.
Chunked prefill~\cite{agrawal2024taming, wu2024loongserve} further improves it to prefill long context in chunks, mitigating execution bubbles, while prefill-decoding disaggregation~\cite{zhong2024distserve} isolates two inference phases, eliminating interference. 
Paged attention~\cite{kwon2023efficient} leverages paged management to eliminate memory fragmentation, while prefix caching~\cite{zheng2024sglang, lin2024parrot} reuses KV cache across requests to further reduce prefilling cost.

\textbf{KV cache management} becomes increasingly critical as the context length increases. 
Beyond the above de facto paged memory, many studies~\cite{yu2025stateful, gao2025fast, jeong2025accelerating, gao2024cost} store KV cache outside the GPU, leveraging host memory and disks.
They swap the inactive KV cache out and swap it back into the GPU when needed.
Some other methods~\cite{zhang2023h2o, xiao2024streamingllm, tang2024quest} are much more aggressive, identifying the cold-spot KV cache and directly evicting it permanently.
\name is orthogonal to these methods that require maintaining tensor formats for GPU direct use, but encodes KV caches into bitstreams to enable efficient cross-node transmission. 

\textbf{Distributed KV cache management} further extends caching capabilities across distributed nodes, to overcome local memory constraints. 
Beyond the systems mentioned in the background, some work~\cite{hu2024memserve, lin2024infinite, fu2024serverlessllm} proposes granular paging and prefetching mechanisms to speed up or mask retrieval delays.
To tackle the transmission issue, some architectural works~\cite{qin2025mooncake,licker2025rdma} explore disaggregated memory pooling via RDMA, uses zero-copy to bypass CPU during data transfer.
Other work~\cite{liu2024cachegen, xiang2025shadowserve} compresses the KV cache before transmission, whereas \name leverages hardware video codecs to enable efficient remote reuse of the KV cache.



\section{Conclusion}
In this paper, we study the remote KV cache reuse for LLM inference. 
We find that existing methods provide only suboptimal KV compression and their decompression either competes for inference resources or requires new hardware devices at extra cost.
Unlike them, we leverages idle GPU-native chips and propose \name system.
\name contains a codec-friendly tensor layout that compress KV caches to compact video formats, and an efficient remote KV fetcher that hides network latency and eliminates resource contention.
In our evaluation across heterogeneous models and hardware, \name delivers significant speedups in TTFT compared to SOTA methods while maintaining high accuracy.
\noindent


\bibliographystyle{plain}
\bibliography{reference}
\appendix

\section{Appendix}\label{sec:appendix}




\subsection{Additional results of image similarity}\label{sec:similarityadditional}
\begin{figure}[h]
     \centering
     \setlength{\abovecaptionskip}{5pt}
    \includegraphics[width=0.6\linewidth]{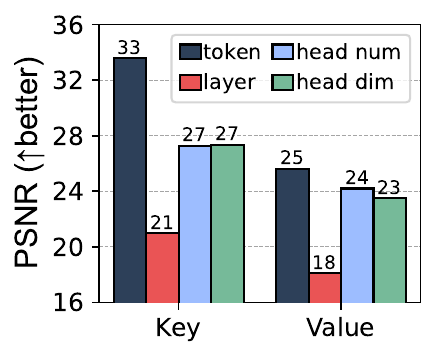}
    \caption{PSNR analysis of slicing KV cache along different dimensions.}
    \label{fig:psnr}
    \vspace{-0.5em}
\end{figure}
Fig.~\ref{fig:psnr} plots the PSNR value of slicing the KV cache along different dimensions.
The same as SSIM in Fig.\ref{fig:ssim}, slicing the KV cache along token dimensions brings the most image similarity.

\begin{figure}[h]
     \centering
    \includegraphics[width=1\linewidth]{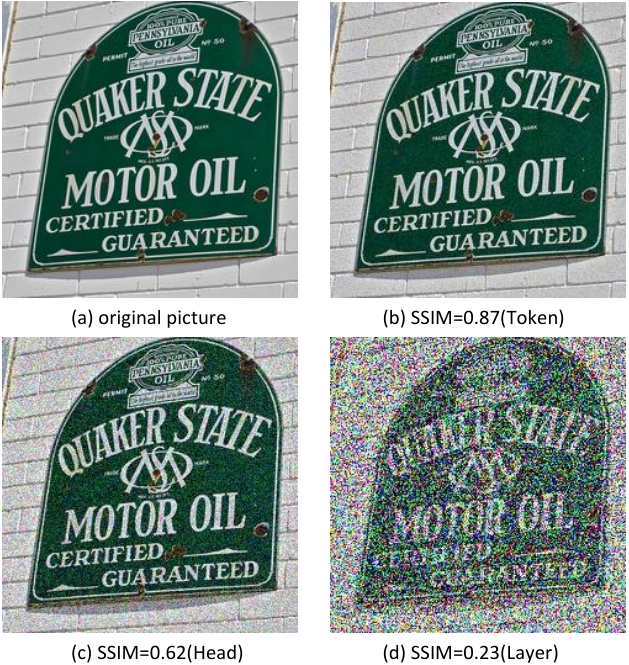}
    \caption{Visualization of structural similarity (SSIM) across different dimensions.}
    \label{fig:ssim_pic}
    \vspace{-0.5em}
\end{figure}

Fig.~\ref{fig:ssim_pic} visualizes the different image similarity levels.
We add different levels of noise to the original picture, leading to the noised picture having a corresponding SSIM with the original picture.
These SSIM values match the SSIM of continuous tensors sliced from the KV cache along different dimensions, \eg, the 0.87 of the token dimension as shown in Fig.\ref{fig:ssim}.


\subsection{Algorithm of adaptive resolution and corresponding lookup table}\label{sec:pseudocode}
\noindent

\begin{algorithm}[h]
\footnotesize 
\caption{Adaptive Resolution Selection via Bubble Minimization}
\label{alg:bubble_min}
\begin{algorithmic}[1]
\Require Historical Bandwidth $\mathcal{B}_{t-1}$, Current Decompress Pool Load $\mathcal{L}_{pool}$, Profile Table $\mathcal{T}_{prof}$, Support resolution set $\mathcal{R}_{all}$
\Ensure Optimal Resolution $r_{opt}$

\State $\hat{B}_{t} \gets$ \Call{EstBandwidth}{$\mathcal{B}_{t-1}$} \Comment{Predict current BW from history}
\State $\delta_{min} \gets \infty$
\State $r_{opt} \gets \text{NULL}$

\For{$r$ \textbf{in} $\mathcal{R}_{all}$} \Comment{Iterate through all candidate resolutions}
    \State $S_r \gets$ \Call{GetVideoSize}{$r$}
    \State $\tau_{trans} \gets S_r / \hat{B}_{t}$ \Comment{Calculate transmission latency}
    \State $\tau_{dec}, \tau_{penalty} \gets$ \Call{LookupTable}{$\mathcal{T}_{prof}, r, \mathcal{L}_{pool}$} \Comment{Get decoding latency with switching penalty}
    
    \State $\delta_{bubble} \gets |\tau_{trans} - \tau_{dec}-\tau_{penalty}|$ \Comment{Calculate the bubble gap}
    
    \If{$\delta_{bubble} < \delta_{min}$}
        \State $\delta_{min} \gets \delta_{bubble}$
        \State $r_{opt} \gets r$
    \EndIf
\EndFor

\State \Return $r_{opt}$

\end{algorithmic}
\end{algorithm}

Adaptive resolution fetching algorithm is shown in Alg.~\ref{alg:bubble_min}, which aims to minimize pipeline bubbles by synchronizing transmission and decoding stages. 
Initially, the available bandwidth $\hat{B}_{t}$ is predicted using historical traces $\mathcal{B}_{t-1}$ (line\#1). 
Next, the algorithm iterates through the set of supported resolutions $\mathcal{R}_{all}$ to evaluate each candidate $r$ (line\#4). 
In each iteration, the transmission latency $\tau_{trans}$ is estimated from the predicted bandwidth and video size $S_r$ (line\#5-6). 
Regarding the decoding latency $\tau_{dec}$, we query the lookup table $\mathcal{T}_{prof}$ based on the current pool load $\mathcal{L}_{pool}$ (line\#7).
The lookup table across different devices is shown in Tab.~\ref{tab:latency_comparison_booktabs}, Tab.~\ref{tab:latency_comparison_booktabs_l20} and Tab.~\ref{tab:latency_comparison_booktabs_l20}.
Furthermore, we observe that resolution switching within the pool affects the processing efficiency of lower-resolution videos.
In addition, to account for resolution switch overheads, a switching penalty $\tau_{penalty}$ whenever the candidate resolution differs from the active resolution in the decompression pool. 
Finally, the system identifies the optimal resolution $r_{opt}$ that minimizes the bubble $\delta_{bubble}$ between transmission and decompression  (line\#8-11).

\begin{table}[h]
\centering
\footnotesize
\begin{tabular}{ccccc}
\toprule
\multirow{2}{*}{Concurrency} & \multicolumn{4}{c}{Latency (s)} \\ 
\cmidrule(lr){2-5}
 & 240P & 480P & 640P & 1080P \\ 
\midrule
1 & 0.21 & 0.2 & 0.2 & 0.19 \\
2 & 0.22 & 0.22 & 0.21 & 0.19 \\
3 & 0.29 & 0.30 & 0.29 & 0.26 \\
4 & 0.32 & 0.31 & 0.30 & 0.30 \\
5 & 0.46 & 0.42 & 0.37 & 0.35 \\
6 & 0.52 & 0.43 & 0.41 & 0.40 \\
7 & 0.62 & 0.51 & 0.45 & 0.43 \\
\midrule 
Penalty & 0.08 & 0.06 & 0.03 & 0 \\ 
\midrule 
Size (MB) & 180 & 205 & 235 & 256 \\ 
\bottomrule
\end{tabular}
\caption{Lookup table of different resolutions on H20.}
\label{tab:latency_comparison_booktabs}
\end{table}

\begin{table}[h]
\centering

\footnotesize
\begin{tabular}{ccccc}
\toprule
\multirow{2}{*}{Concurrency} & \multicolumn{4}{c}{Latency (s)} \\ 
\cmidrule(lr){2-5}
 & 240P & 480P & 640P & 1080P \\ 
\midrule
1 & 0.18 & 0.175 & 0.17 & 0.16 \\
2 & 0.18 & 0.178 & 0.175 & 0.16 \\
3 & 0.19 & 0.183 & 0.175 & 0.161 \\
\midrule 
Penalty & 0.06 & 0.06 & 0.04 & 0 \\ 
\midrule 
Size (MB) & 180 & 205 & 235 & 256 \\ 
\bottomrule
\end{tabular}
\caption{Lookup table of different resolutions on L20.}
\label{tab:latency_comparison_booktabs_l20}
\end{table}

\begin{table}[h]
\centering

\footnotesize
\begin{tabular}{ccccc}
\toprule
\multirow{2}{*}{Concurrency} & \multicolumn{4}{c}{Latency (s)} \\ 
\cmidrule(lr){2-5}
 & 240P & 480P & 640P & 1080P \\ 
\midrule
1 & 0.25 & 0.24 & 0.231 & 0.20 \\
2 & 0.252 & 0.241 & 0.235 & 0.21 \\
3 & 0.252 & 0.25 & 0.24 & 0.22 \\
4 & 0.26 & 0.26 & 0.25 & 0.24 \\
5 & 0.29 & 0.27 & 0.27 & 0.25 \\
\midrule 
Penalty & 0.04 & 0.04 & 0.03 & 0 \\ 
\midrule 
Size (MB) & 180 & 205 & 235 & 256 \\ 
\bottomrule
\end{tabular}
\caption{Lookup table of different resolutions on A100.}
\label{tab:latency_comparison_booktabs_a100}
\end{table}

\subsection{Layer-wise fetching-inference pipeline.}\label{sec:layerwisepipeline}

\begin{figure}[t]
     \centering
    \includegraphics[width=0.9\linewidth]{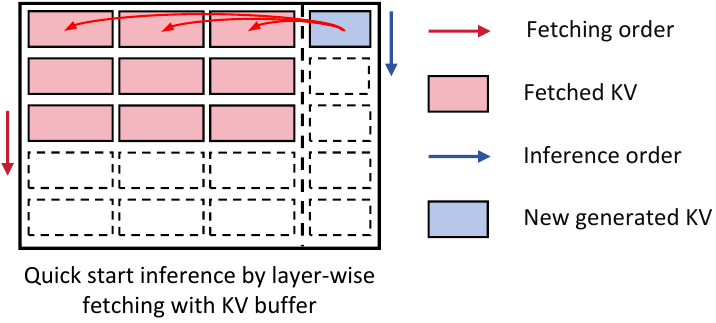}
    \caption{Layer-wise KV fetching with KV buffer.}
    \label{fig:layerwise}
\end{figure}

We propose a layer-wise KV fetching pipeline with a KV buffer to achieve a non-blocking execution pipeline as shown in Fig.~\ref{fig:layerwise}.
Following layer-wise pipeline design in Mooncake, \name also pre-allocates GPU memory that entire KV cache required, and fill the fetched remote KV cache into GPU memory in a layer-by-layer manner.
To prevent fetching from stalling non-reuse requests, we maintain a KV buffer to track the state of each layer's KV cache. 
Fetching requests are added to the running queue only when the pipeline satisfies following non-blocking condition:
$$\sum_{j=1}^{k} T_{\text{decode}}^{(j)} \le \sum_{j=1}^{k-1} T_{\text{comp}}^{(j)}, \quad \forall k \in \{L_{\text{buf}} + 1, \dots, L_{\text{total}}\}$$
where $T_{\text{decode}}^{(j)}$ and $T_{\text{comp}}^{(j)}$ denote the decoding and computation time for the $j$-th layer, $L_{\text{buf}}$ and $L_{\text{total}}$ denote the buffered layer number and total layer number.
This ensures that for every unbuffered layer $k$, the KV cache data is ready exactly before the GPU completes the computation of layer $k-1$, thereby preventing any execution stalls.
Thanks to the de facto chunked prefill and sequence parallelism techniques, the computation time of each layer can be predicted very precisely.

\end{document}